\RequirePackage[T1]{fontenc}
\IfFileExists{IEEEtran.cls}{\documentclass[journal]{IEEEtran}}{\documentclass[10pt,twocolumn]{article}\usepackage[margin=19mm]{geometry}}
\usepackage{amsmath,amssymb,booktabs,array}
\usepackage{graphicx,placeins}
\graphicspath{{figures/}}

\usepackage[hidelinks]{hyperref}
\newcommand{\DI}{\mathrm{DI}}
\newcommand{\eff}{\mathrm{eff}}
\newcommand{\RMSE}{\mathrm{RMSE}}
\newcommand{\dd}{\mathrm{d}}
\newcommand{\tr}{^{\mathsf T}}

\title{Spatial-Mode Selection for Local Mineral Absorption Spectroscopy}

\author{Qipeng~Qian, Jialiang~Jin, Shihao~Wang, and~Yuntao~Qian%
\thanks{Qipeng Qian is with the College of Artificial Intelligence, Zhejiang University, Hangzhou 310027, China, and also with SUPCON Technology, Hangzhou, China (e-mail: 0626c67@zju.edu.cn).}%
\thanks{Jialiang Jin is with the Department of Computer Science, Whiting School of Engineering, Johns Hopkins University, Baltimore, MD 21218, USA (e-mail: jjin45@jh.edu).}%
\thanks{Shihao Wang is with the Department of Mathematics, University of Wisconsin-Madison, Madison, WI 53706, USA (e-mail: swang2232@wisc.edu).}%
\thanks{Yuntao Qian is with the College of Artificial Intelligence, Zhejiang University, Hangzhou 310027, China (e-mail: ytqian@zju.edu.cn).}%
}
\date{}
\begin{document}
\maketitle

\begin{abstract}
Absorption-band depth measures the fractional decrease in mineral reflectance below its local continuum. Recovering this feature strength for a small mineral region supports quantitative comparisons between regions, but diffraction mixes its light with the surroundings. Conventional spectral fitting can separate mean contributions; it cannot remove background photon noise already recorded. We develop a quantum-inspired measurement framework that uses spatial-mode demultiplexing (SPADE) to retain the target absorption response while reducing background acceptance before detection. The framework connects finite mineral-region geometry and reference spectra to estimation precision. A loss-dependent design condition identifies when adding SPADE exposure improves on direct imaging (DI) under fixed-response Poisson measurements with unknown scene parameters. Joint inference from initial DI observations and both mode ports estimates target depth together with an unknown, spatially shared background spectrum. In equal-time simulations that include localization uncertainty, alignment time, calibration, and detector costs, this acquisition reduces depth root-mean-square error (RMSE) by 14\%--26\% across five absorption strengths relative to all-pixel DI with the same background freedom. These results use a supplied target feature shape and geometry, with target continuum slope fitted and curvature fixed. The analysis separates optical selectivity from parameter ambiguity and acquisition cost, providing a quantitative basis for spatial-mode receiver design in local mineral spectroscopy.

\end{abstract}

\section{Introduction}\label{sec:intro}
Mineral imaging spectroscopy records pixel reflectances to identify minerals and compare rocks or outcrops \cite{goetz1985,kruse,greenberger,vandermeer2012}. After selecting a grain, vein, or altered patch, a quantitative question remains: how strong is an absorption feature in that region's own spectrum? Answering it requires a feature-strength measure and separation of target light from its surroundings.

\begin{figure*}[!t]\centering
\includegraphics[width=\textwidth]{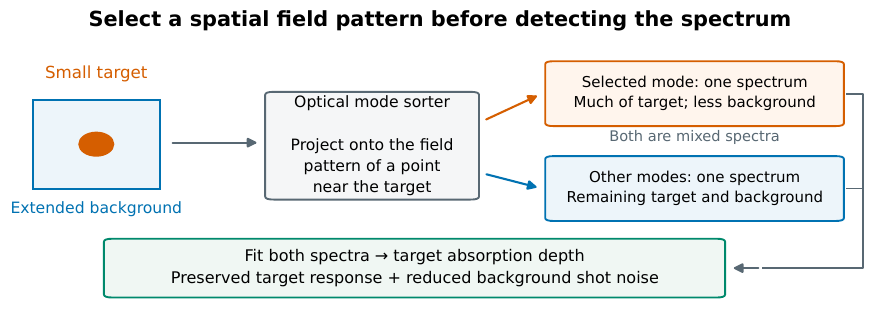}
\caption{Binary SPADE separates a selected optical field mode from its orthogonal complement and records a spectrum at each port. A mode matching the field from a point near the target can retain much of a small target's light while directing much of an extended background to the other port. Both spectra remain target/background mixtures and are fitted jointly for depth. DI records pixel spectra instead. The schematic illustrates the principle; precision depends on scene geometry, spectra, and receiver loss.}\label{fig:research_overview}
\end{figure*}

An absorption feature is a reflectance dip. Its \emph{continuum} is a smooth baseline fitted across the band from its shoulders. Depth is the dip below that baseline divided by the baseline value \cite{clarkroush}:
\[
\text{Band depth}
=\frac{\text{continuum reflectance}-\text{band reflectance}}
{\text{continuum reflectance}}.
\]
This dimensionless spectral depth describes the strength of the dip; it is distinct from light penetration depth or mineral abundance.

Band position helps identify the absorption mechanism; depth measures strength and can vary between regions with the same band position \cite{hunt1977,clark1990}. Greenberger \emph{et al.} map silica/clay variations between sedimentary layers and ferric-iron variations in serpentinized rock \cite{greenberger}. Liu \emph{et al.} relate carbonate feature strengths and depth ratios to independently measured mineral content \cite{liu2023}. Such uses require material-specific interpretation.

Continuum normalization removes common brightness scaling, but grain size, scattering, and mixtures still affect depth \cite{clarkroush,hapke1981}. Its uncertainty governs the reliability of spectral-contrast comparisons. EMIT Level 2B reports depths and uncertainties with mineral identities \cite{emitdepth}. We address depth precision for a selected local target overlapping its surroundings.

A resolved, low-noise target permits direct depth measurement. For a small target, diffraction mixes its light with background in several pixels. Flat background dilutes the dip; structured background can change its shape. Mixed-pixel depth therefore need not equal target depth, whose recovery requires wavelength-dependent contributions from both materials.

Continuum removal and feature fitting estimate depth \cite{clarkroush,asadzadeh2016}; mixing models relate mixed pixels to reference spectra and contributions \cite{manolakis}. Even exact mean-background removal leaves background photon fluctuations. Precision therefore also depends on the optical collection preceding spectral inference.

Direct imaging (DI) records pixel spectra. Selecting nearby pixels reduces background but discards diffracted target light. All-pixel fitting retains information and models mean mixtures, yet cannot remove recorded background shot noise. We use this full DI baseline and ask whether another optical measurement preserves the target response with less accompanying background noise.

Quantum estimation theory motivates spatial-mode measurements, which outperform DI for incoherent-source separation \cite{tsang,nair2016} and have theoretical/experimental spectroscopy precedents \cite{huang,santamaria}. They act on the optical field without requiring entanglement. We extend them to depth estimation for a finite reflecting mineral patch in an absorbing background.

A spatial mode is an amplitude/phase pattern. Binary spatial-mode demultiplexing (SPADE) separates the field matching a selected pattern from its orthogonal complement and records two spectra (Fig.~\ref{fig:research_overview}). Choosing a point-like field near the target preserves much of a small target's light while directing much of an extended background elsewhere. Preserved target response and reduced shot noise can improve depth precision. Both ports remain mixtures and are fitted jointly; the target's physical depth is unchanged.

A small error in bright-background spectra can mimic weak-target absorption. We therefore learn the background with depth from the same observations: initial DI supplies localization and spatial background information, while the mode ports add complementary target/background mixtures.

This follow-up measurement assumes supplied target feature shape/geometry and background brightness distribution. Depth, brightness, and exact position are unknown. We test supplied spectral shapes, then a freely estimated spatially shared background spectrum. Both receivers share aperture, illumination, spectral response, scene-parameter freedom, and total time. Adaptive DI localization \cite{grace} incurs uncertainty and alignment/calibration time; mode loss is included and commands are assumed accurately executed.

The work makes three contributions:
\begin{enumerate}
\item \emph{A finite-region model for local mineral spectroscopy.} Measured reference spectra, finite target geometry, diffraction, and mode projection are connected to absorption-depth information. The model explains the benefit of optical selection relative to all-pixel DI, including optimal local pixel weighting.
\item \emph{A receiver-specific design condition with loss and nuisance coupling.} We specialize local optimal-design theory to DI--SPADE acquisition and obtain a necessary and sufficient condition for a beneficial SPADE exposure under fixed-response Poisson measurements. The condition identifies the transmission needed to compensate for displaced DI exposure and unknown scene parameters.
\item \emph{Joint inference of depth and a spatially shared unknown background spectrum.} Initial DI data and both mode ports jointly constrain target depth and one positive background spectral value per model wavelength, shared across the background region. Complete-acquisition simulations compare this strategy with equally corrected DI under a common time budget; separate material and mismatch tests identify how spectral and spatial structure affect the gain.
\end{enumerate}

Section~\ref{sec:related} reviews prior work; Secs.~\ref{sec:background}--\ref{sec:information} develop the model and design criterion; Secs.~\ref{sec:experiments}--\ref{sec:conclusion} give experiments and conclusions. Appendices retain derivations, implementation, and sensitivity tests.

\section{Related Work}\label{sec:related}
Prior work addresses mineral-feature extraction, optical measurement, and observation design with unknown position.

\subsection{Mineral spectroscopy and system modeling}
Continuum removal defines depth relative to a local baseline \cite{clarkroush}; band-decomposition models also separate overlapping absorptions \cite{sunshine1990}. Asadzadeh and de Souza Filho fit sampled, noisy continuum-removed features iteratively with Gaussian/polynomial variants \cite{asadzadeh2016}. Liu \emph{et al.} connect carbonate strengths/ratios to independent mineralogy and show restrictions from overlapping features \cite{liu2023}.

Reference-spectrum mixing models support subpixel detection and mismatch analysis \cite{manolakis,bioucas2012}. Our target is already selected and its continuous depth is unknown. Feature fitting infers parameters from recorded spectra; optical selection instead changes the photon mixtures recorded.

Swayze \emph{et al.} examine bandpass, sampling, and noise in spectral identification \cite{swayze}; Kerekes and Baum connect scene/system models to subpixel detection and airborne measurements \cite{kerekes}. We adopt this forward-modeling approach, using measured USGS reflectances \cite{usgs}, receiver-specific spatial probabilities, and analytical verification.

\begin{figure*}[!t]\centering
\includegraphics[width=\textwidth]{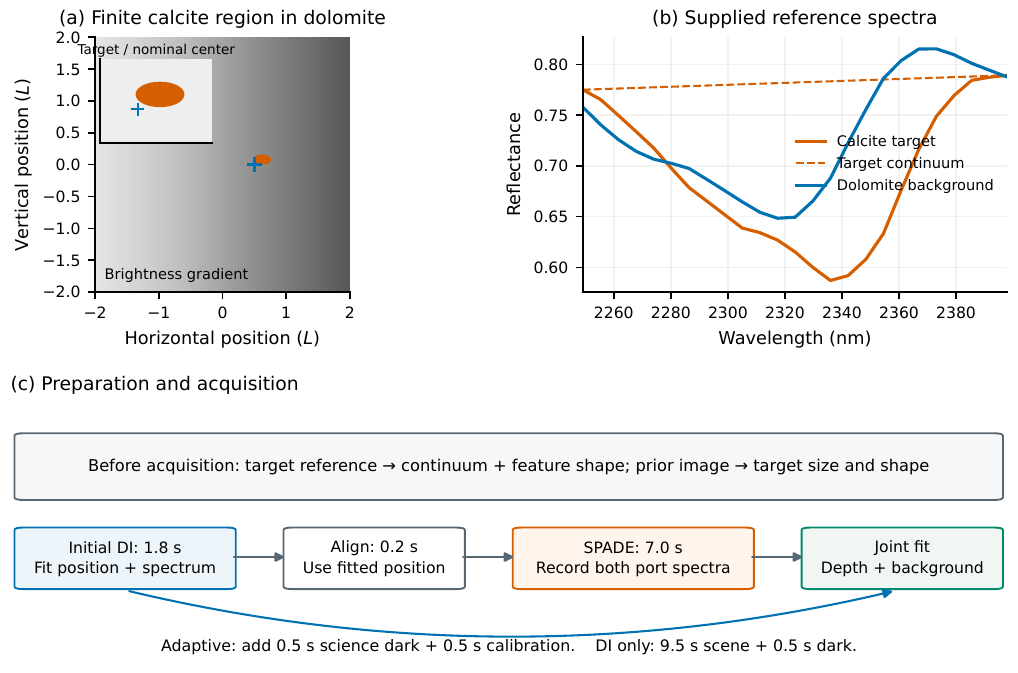}
\caption{C4 scene and acquisition. (a) Finite calcite target in dolomite with a left-to-right brightness gradient, before diffraction. The blue cross is the fixed-comparison mode center; the inset shows its offset from the target. Coordinates use $L=1.09975$ mm [(\ref{eq:spatial_scale})]. (b) USGS component reflectances before mixing/filtering. Target depth is the fractional dip below the orange dashed continuum at maximum absorption. (c) Preparation supplies target shape and geometry. DI observes for 9.5 s; adaptive acquisition uses 1.8 s DI localization, 0.2 s alignment, and 7.0 s dual-port observation, fitted jointly. Dark and calibration complete the 10 s budgets (Table~\ref{tab:time}).}\label{fig:overview}
\end{figure*}

\subsection{Spatial-mode measurements and spectroscopy}
Tsang, Nair, and Lu establish spatial-mode advantages for incoherent-source separation \cite{tsang}. Spatial-mode and phase-sensitive experiments demonstrate sub-Rayleigh source measurements \cite{paur2016,tham2017,tang2016,tan2023}. Boucher \emph{et al.} demonstrate calibrated two-dimensional distance estimation \cite{boucher2020}; Wallis \emph{et al.} jointly estimate separation and relative power \cite{wallis2025}. Ansari \emph{et al.} demonstrate related temporal-mode timing/intensity measurements \cite{ansari}.

Huang \emph{et al.} analyze absorption-line discrimination \cite{huang}; Santamaria \emph{et al.} demonstrate mode-resolved spectroscopy and already formulate continuous weak-source spectral estimation \cite{santamaria}. We benchmark their two-source probabilities/scaling. Our increment is finite reflecting targets, extended absorbing backgrounds, and joint depth, brightness, position, and shared-spectrum estimation under acquisition costs.

\subsection{Alignment and observation design}
Grace \emph{et al.} establish DI localization followed by binary mode detection \cite{grace}. Prasad motivates excluding the extended background from target support \cite{prasad}. We apply both ideas to reflecting surfaces, retaining initial DI in the final fit and charging alignment/calibration to the common budget. Unknown relative brightness can also alter phase-sensitive precision gains \cite{bonsma2019}, motivating joint nuisance-parameter inference.

Classical $c$-optimal design minimizes a selected parameter's local variance bound \cite{elfving1952,pukelsheim}. We specialize its directional condition to absorption under lossy DI--SPADE and nuisance coupling. The increment is this receiver-specific mineral design and unknown-background fit; the two-stage architecture remains that of Grace \emph{et al.}

\section{Measurement Model}\label{sec:background}
The forward model defines target depth, integrates receiver collection over the finite scene, and predicts noisy counts for inference. Both receivers share the scene, aperture, and illumination. Figure~\ref{fig:overview} shows the pre-diffraction scene, component reflectances, and acquisition: an extended background may contribute far more photons than the small target despite comparable reflectances. Section~\ref{sec:setup} gives numerical settings.

\subsection{Defining local absorption depth}\label{sec:parameter}
We estimate one band's depth in the target spectrum before optical mixing. Its reference-supplied position, width, and normalized shape remain fixed; scaling depth also scales continuum-normalized band area.

At wavelength $\lambda_\ell$, subscripts $T,B$ denote target/background and superscript $\mathrm{ref}$ a supplied reference. Let $C_\ell>0$ be the smooth continuum fitted to the shoulders of target reference $R^{\rm ref}_{T\ell}$. Both are dimensionless reflectances; the continuum is a fitted baseline, not a separate absorption-free measurement. Normalizing the fractional dip gives

\begin{equation}\label{eq:profile_definition}
h_\ell=\frac{1-R^{\rm ref}_{T\ell}/C_\ell}
{\displaystyle\max_{\ell'}(1-R^{\rm ref}_{T\ell'}/C_{\ell'})}.
\end{equation}
The numerator is the local fractional dip and the denominator its nonzero band maximum, with $\ell'$ running over the band. Thus $h_\ell=1$ at maximum absorption and zero where there is no dip, independently of reference depth.

At reference brightness, the unknown depth $\alpha$ defines
\begin{equation}\label{eq:feature_background}
R_{T\ell}(\alpha)=C_\ell(1-\alpha h_\ell),
\qquad \max_\ell h_\ell=1.
\end{equation}
At $h_\ell=1$, $\alpha$ is the fractional continuum decrease; increasing it deepens the fixed-shape band and leaves $h_\ell=0$ wavelengths unchanged. Errors are dimensionless depth errors. Later spectral filtering changes predicted counts, not the parameter definition.

Both receivers receive the same shape, target geometry, background spatial model, and approximate position; depth, brightnesses, and exact position are inferred. We first supply correct spectral shapes, then estimate the background freely (Sec.~\ref{sec:background_learning}) and test reference perturbations (Sec.~\ref{sec:joint_results}).

Unknown factors $\beta_T,\beta_B$ scale the entire target/background spectra, with unity retaining reference brightness:
\begin{equation}\label{eq:spectrum}
R_{T\ell}=\beta_T C_\ell(1-\alpha h_\ell),
\end{equation}
\begin{equation}\label{eq:background_spectrum}
R_{B\ell}=\beta_B R^{\rm ref}_{B\ell}.
\end{equation}
Here $R^{\rm ref}_{B\ell}$ includes background absorption. Unlike brightness, depth scales only the target dip. The unknown-background model replaces $\beta_B$ by free spectral values and additionally fits target continuum slope (Appendix~\ref{app:references}).

Their spectral responses differ:
\begin{equation}\label{eq:spectral_sensitivities}
\begin{aligned}
\partial_\alpha R_{T\ell}&=-\beta_T C_\ell h_\ell,\\
\partial_{\beta_T}R_{T\ell}&=C_\ell(1-\alpha h_\ell).
\end{aligned}
\end{equation}
Depth changes wavelengths with large $h_\ell$, whereas brightness also scales the continuum. Band and shoulder measurements can distinguish them; a single wavelength, or constant $h_\ell$ across all channels, reveals only $\beta_T(1-\alpha h_\ell)$ and cannot separate the two. Diffraction, filtering, and background mixing may create further parameter ambiguities, quantified in Sec.~\ref{sec:information}.

\subsection{Direct imaging and spatial-mode projection}\label{sec:receivers}
DI records a spectrum per image pixel. Binary SPADE records two spectra: one for a selected field mode and one for its entire orthogonal complement. ``Binary'' denotes spatial ports, not wavelength channels.

For range $z$, aperture diameter $D$, and fixed reference wavelength $\lambda_{\rm ref}$, define the object-plane diffraction length
\begin{equation}\label{eq:spatial_scale}
L=\frac{\lambda_{\rm ref}z}{D}.
\end{equation}
All coordinates are divided by $L$; image coordinates are first mapped to the object plane by magnification. Thus coordinate $0.5$ represents $0.5L$. This reference scale grows with range and decreases with aperture; it is neither pixel pitch nor an exact resolution cutoff.

Let $\mathbf u=(u_x,u_y)$ be a source point and $\mathbf x$ an image coordinate in these units. The normalized amplitude point-spread function (PSF) $\psi_{\ell,\mathbf u}(\mathbf x)$ is its image field at wavelength $\lambda_\ell$. Its squared magnitude is the spatial probability density for photon detection, integrating to one over the image plane. DI integrates this density over a pixel.

We select the PSF field $\psi_{\ell,\mathbf u_0}$ centered on the commanded position $\mathbf u_0$, which may differ from the true target center. The ideal probability of a photon from $\mathbf u$ entering selected port $0$ is
\begin{equation}\label{eq:projection_background}
p_0(\mathbf u,\ell)=\left|
\int\psi_{\ell,\mathbf u_0}^{*}(\mathbf x)
\psi_{\ell,\mathbf u}(\mathbf x)\dd^2\mathbf x\right|^2.
\end{equation}
Here $*$ denotes conjugation and $\dd^2\mathbf x$ image-plane area. The integral adds normalized field amplitudes, including phase, before squaring. Matching fields give probability one; orthogonal fields give zero even if their intensities spatially overlap. This is field similarity, not overlapping intensity area. Later we include loss and port crosstalk.

DI weights combine already recorded counts; mode projection combines optical amplitudes before photon counting. Pixel weights cannot reproduce this phase-sensitive operation. Both use the same aperture/PSF: SPADE changes depth-estimation precision, not PSF width or spectral resolution.

Distinct surface points are mutually incoherent, so we project each point's field and sum its intensity-weighted probability. We never coherently add target/background fields. This spatially incoherent model excludes coherent speckle.

For a circular pupil the selected field is an Airy amplitude PSF. A small target near $\mathbf u_0$ can couple strongly, while an extended background has lower average coupling; nearby background still enters. The sorter selects field pattern and position, not mineral identity.

The complement collects probability $1-p_0$, covering all orthogonal modes rather than only first order. Both ports help estimate background and other unknowns. Initial DI estimates position and sets $\mathbf u_0$.

Quantum estimation theory motivates selecting optical measurements for a parameter \cite{tsang}, although these probabilities also follow from classical wave optics. We compare classical Fisher information for DI and this specified receiver, without computing mineral-scene quantum Fisher information or claiming quantum optimality.

Finite target and background count rates follow by integrating these point-source probabilities.

\subsection{From a reflecting patch to photon counts}\label{sec:model}
The binary mask $m_T(\mathbf u)$ defines the finite target. The nonnegative background weight $m_B(\mathbf u)$ is zero within it and outside the source field, avoiding double counting. Unlike $m_T$, it may exceed one: doubling it doubles local background contribution. Both materials are Lambertian diffuse reflectors with mutually incoherent surface points. Spectral irradiance is constant over the modeled window; each scene specifies geometry, brightness weight, and reference shapes.

For material $c\in\{T,B\}$, integrate over DI pixel region $\mathcal P_j$ or SPADE ports $j=0,\perp$. The resulting collection coefficients are
\begin{equation}\label{eq:di}
A^{(\DI)}_{c\ell j}=\int m_c(\mathbf u)
\int_{\mathcal P_j}|\psi_{\ell,\mathbf u}(\mathbf x)|^2
\dd^2\mathbf x\,\dd^2\mathbf u,
\end{equation}
\begin{equation}\label{eq:mode}
\begin{aligned}
A^{(\mathrm{SPADE})}_{c\ell0}&=\int m_c(\mathbf u)
p_0(\mathbf u,\ell)\dd^2\mathbf u,\\
A^{(\mathrm{SPADE})}_{c\ell\perp}&=A_c-A^{(\mathrm{SPADE})}_{c\ell0},\\
A_c&=\int m_c(\mathbf u)\dd^2\mathbf u .
\end{aligned}
\end{equation}
We omit receiver superscripts when a formula applies to either. Each $A_{c\ell j}$ is material area weighted by its probability of reaching output $j$; e.g., $A^{(\mathrm{SPADE})}_{T\ell0}$ collects target light in the selected port. These are normalized object-plane effective areas, not detector/aperture areas. Their unselected total is $A_c$; ideal mode coefficients sum to it, while finite DI coverage leaves unrecorded light.

Let $\tau$ be exposure excluding readout and $K_\ell$ the photoelectron rate per unit reflectance and normalized area. Summing incoherent target/background rates gives
\begin{equation}\label{eq:mean_main}
\mu_{\ell j}=\tau K_\ell\sum_{c\in\{T,B\}}R_{c\ell}A_{c\ell j}.
\end{equation}
The repeated-exposure mean $\mu_{\ell j}$ uses identical reflectances but receiver-specific areas. Appendix~\ref{app:optics} derives $K_\ell$ from irradiance, range/aperture, photon energy, and common efficiency. Spectral filtering, transmission, and crosstalk precede detector shot, dark, and read noise (Appendix~\ref{app:noise}).

Absorption changes target counts; background adds noise without directly responding to $\alpha$. We next learn its spectrum from the same observations.

\subsection{Learning the background spectrum during acquisition}\label{sec:background_learning}
Instead of fixing background shape and fitting brightness, we estimate a positive background value at each of 25 model wavelengths, before spectral filtering:
\begin{equation}\label{eq:free_background}
R_{B\ell}=R^{\rm ref}_{B\ell}\exp(\gamma_\ell),
\qquad \ell=1,\ldots,25.
\end{equation}
The logarithmic correction $\gamma_\ell$ leaves the reference unchanged at zero and increases/decreases it with its sign. Positivity is enforced by the exponential. This replaces (\ref{eq:background_spectrum}) and includes overall background brightness, removing a separate $\beta_B$.

All 25 corrections are free without smoothing. The reference is a starting coordinate system for positive spectra; replacement shifts the corrections but can leave the physical fitted spectrum unchanged. Appendix~\ref{app:reference_invariance} gives the transformation and numerical-bound conditions. Such replacement tests fitting stability; target-template or spatial-background changes test different physical assumptions.

Target and background changes follow different DI patterns, $A^{(\DI)}_{T\ell j}$ and $A^{(\DI)}_{B\ell j}$. The mode ports supply two further mixtures: the selected port preserves target response, while the complement constrains background. At one wavelength, one port cannot distinguish absorption from freely varying background; a second spatial mixture can. Across wavelengths, joint fitting also separates depth, brightness, and target continuum slope, conditional on the supplied masks/response.

Initial DI jointly estimates position, depth, target brightness/slope, and background; its center estimate commands SPADE. The final fit combines those initial observations once, both ports, and calibration. Background is updated jointly, propagating its uncertainty to depth. Pure DI fits the same free spectrum using all pixels and its allocated exposure. Appendix~\ref{app:noise} gives the command-conditional likelihood.

\section{Fisher Information Analysis}\label{sec:information}
Fisher information compares the count response to absorption against noise, and the Cram\'er--Rao bound converts it to local precision. We proceed from depth alone to joint unknowns and exposure allocation, then test optical selection and the design criterion (Secs.~\ref{sec:ideal_results}--\ref{sec:allocation_results}) before complete acquisitions.

\subsection{Fisher information for absorption depth}

First suppose brightness, position, and background are known. Let independent Poisson count $N_i$ have positive mean/variance $\mu_i(\alpha)$, where $i$ combines wavelength and spatial pixel/port. Depth information is

\begin{equation}\label{eq:scalar_fisher_background}
F_{\alpha\alpha}=\sum_i
\frac{(\partial_\alpha\mu_i)^2}{\mu_i}.
\end{equation}
Here the numerator is squared count sensitivity and the denominator shot-noise variance, summed over all channels. A known mean background still contributes photon noise. Appendix~\ref{app:allocation} derives this result.

Under standard regularity conditions, a locally unbiased depth estimator satisfies
\begin{equation*}
\operatorname{Var}(\widehat\alpha)\geq\frac{1}{F_{\alpha\alpha}}.
\end{equation*}
Local unbiasedness requires the correct mean and first-order response near the evaluated depth. The standard-deviation bound is $1/\sqrt{F_{\alpha\alpha}}$; Monte Carlo separately measures achieved error, including bias.

To isolate spatial selection, consider wavelength $\lambda_\ell$ before filtering, with all other parameters known.

Define retention fractions $f_{T\ell j},f_{B\ell j}$ as the fraction of each material's collected light entering output $j$, normalized by that material's total, not by the detected mixture. Let $S_\ell$ be the target rate without absorption and $B_\ell$ the background rate including its own features. From the effective collection areas,

\begin{equation}\label{eq:retention_definitions}
\begin{alignedat}{2}
f_{T\ell j}&=\frac{A_{T\ell j}}{A_T},&\qquad
f_{B\ell j}&=\frac{A_{B\ell j}}{A_B},\\
S_\ell&=K_\ell\beta_T C_\ell A_T,&
B_\ell&=K_\ell R_{B\ell}A_B.
\end{alignedat}
\end{equation}
The fractions lie in $[0,1]$. Ideal SPADE fractions sum to one per material, whereas finite-field DI may lose light. They precede additional mode loss; rates include common efficiencies.

For exposure $\tau$, the selected target and background rates give
\begin{equation}\label{eq:retention_mean}
\mu_{\ell j}=\tau\left[S_\ell(1-\alpha h_\ell)f_{T\ell j}
+B_\ell f_{B\ell j}\right],
\end{equation}
With fixed spatial fractions,
\begin{equation*}
\partial_\alpha\mu_{\ell j}=-\tau S_\ell h_\ell f_{T\ell j}.
\end{equation*}
Substitution into (\ref{eq:scalar_fisher_background}) yields the channel contribution
\begin{equation}\label{eq:retention_fisher}
F_{\alpha\alpha}^{(\ell j)}=
\frac{\tau S_\ell^2h_\ell^2 f_{T\ell j}^2}
{S_\ell(1-\alpha h_\ell)f_{T\ell j}+B_\ell f_{B\ell j}}.
\end{equation}
Its numerator is exposure times the squared target count-rate change per unit depth; the denominator is total count rate. When background shot noise dominates, $B_\ell f_{B\ell j}\gg S_\ell(1-\alpha h_\ell)f_{T\ell j}$ with $B_\ell f_{B\ell j}>0$,
\begin{equation}\label{eq:background_limit}
F_{\alpha\alpha}^{(\ell j)}\simeq
\frac{\tau S_\ell^2h_\ell^2}{B_\ell}
\frac{f_{T\ell j}^2}{f_{B\ell j}}.
\end{equation}
Thus information scales as $f_{T\ell j}^2/f_{B\ell j}$: retaining more target light at comparable background preserves greater absorption response without comparable noise. If target shot noise dominates, the exact expression instead scales as $f_{T\ell j}$. Reducing total counts alone does not ensure improvement.

Our DI baseline retains all recorded pixels/wavelengths. With other parameters known, signed numerical weights $w_i$ combine counts with response $\sum_iw_i\partial_\alpha\mu_i$ and variance $\sum_iw_i^2\mu_i$, giving

\begin{equation}\label{eq:weights}
\max_{\mathbf w}\frac{(\sum_i w_i\partial_\alpha\mu_i)^2}
{\sum_i w_i^2\mu_i}=F^{(\DI)}_{\alpha\alpha}.
\end{equation}
Equality at $w_i\propto(\partial_\alpha\mu_i)/\mu_i$ recovers all-pixel DI information, including pixels outside a target mask (Appendix~\ref{app:allocation}). These local weights are not optical probabilities or a globally optimal estimator; processing the same counts cannot exceed their full information. SPADE changes the optical means/noise before detection.

We next account for the scene parameters that must be inferred with depth.

\subsection{Joint estimation with nuisance parameters}
Write $\boldsymbol\theta=(\alpha,\boldsymbol\nu)$, where nuisance parameters $\boldsymbol\nu$ affect data but are not the quantity of interest. The baseline includes brightnesses $\beta_T,\beta_B$ and two center coordinates; relevant tests add continuum, background, and calibration unknowns.

Brightness scales spectra; position changes the target mask and the background excluded by it. Their count responses can resemble absorption. The Poisson information matrix is
\begin{equation}\label{eq:poisson_matrix}
F_{ab}=\sum_i\frac{(\partial_{\theta_a}\mu_i)
(\partial_{\theta_b}\mu_i)}{\mu_i}.
\end{equation}
Indices $a,b$ identify parameters and $i$ all observed channels. Diagonals measure sensitivity; cross terms measure similarity of parameter responses weighted by noise. Partitioning depth and nuisance blocks gives

\begin{equation}\label{eq:efficient}
\begin{aligned}
F&=\begin{pmatrix}
F_{\alpha\alpha}&F_{\alpha\nu}\\
F_{\nu\alpha}&F_{\nu\nu}
\end{pmatrix},\\[3pt]
I_\eff&=F_{\alpha\alpha}
-F_{\alpha\nu}F_{\nu\nu}^{-1}F_{\nu\alpha}.
\end{aligned}
\end{equation}
Effective depth information $I_\eff$ subtracts nuisance coupling from the depth information $F_{\alpha\alpha}$ with nuisance parameters known. For brightness alone the penalty is $F_{\alpha\beta_T}^2/F_{\beta_T\beta_T}$.

For regular locally unbiased joint estimation,
\begin{equation}\label{eq:marginal_bound}
\operatorname{Var}(\widehat\alpha)\geq[F^{-1}]_{\alpha\alpha}
=\frac{1}{I_\eff}.
\end{equation}
This inverts the full matrix before selecting the depth entry; it generally differs from $1/F_{\alpha\alpha}$. The Schur-complement identity requires invertible $F_{\nu\nu}$ and $I_\eff>0$. With free background, 25 coefficients replace $\beta_B$. Numerically, whiten count derivatives by $\sqrt{\mu_i}$ and project depth response away from nuisance responses; its residual squared norm is $I_\eff$. This agrees with the Schur complement and also handles redundant directions (Appendix~\ref{app:allocation}).

This penalty is lost parameter distinguishability, not lost photons. If nuisance changes reproduce a depth change, $I_\eff=0$; near-dependence gives a large variance bound. More photons cannot resolve exact ambiguity, such as depth/brightness at one wavelength. Consequently, DI or the complement can improve depth by constraining position/background even with little direct depth sensitivity; high target counts alone are insufficient.

\subsection{Exposure-time allocation}
Exposure given to SPADE displaces DI. First fix receiver responses (DI sampling and commanded center), use the same unknown parameters, and require an invertible DI nuisance block. Poisson information scales linearly with exposure and known parameter-independent transmission $\eta$; independent measurements add (Appendix~\ref{app:allocation}).

Let $F^{(\DI)},F^{(\mathrm{SPADE})}$ be the two information matrices for the same reference exposure, with SPADE evaluated before additional mode loss. A fraction $t\in[0,1]$ of that exposure goes to SPADE and $1-t$ to DI. Transmission $\eta\in[0,1]$ is additional to their common efficiency:
\begin{equation}\label{eq:allocation}
F(t)=(1-t)F^{(\DI)}+t\eta F^{(\mathrm{SPADE})}.
\end{equation}
Define $I_{\DI}=I_\eff(F^{(\DI)})>0$. A small joint change $\delta\boldsymbol\theta=v\,\delta\alpha$ has $v_\alpha=1$ and nuisance components specifying compensating changes per unit depth.

Minimizing $v\tr F^{(\DI)}v$ over nuisance components gives DI's least distinguishable depth--nuisance direction:

\begin{equation}\label{eq:di_direction}
\begin{aligned}
v_{\DI}&=\begin{pmatrix}
1\\-[F_{\nu\nu}^{(\DI)}]^{-1}F_{\nu\alpha}^{(\DI)}
\end{pmatrix},\\[3pt]
v_{\DI}\tr F^{(\DI)}v_{\DI}&=I_{\DI}.
\end{aligned}
\end{equation}
This is an infinitesimal direction, not a finite unit depth change; e.g., its brightness component describes how brightness partly conceals absorption.

\emph{Criterion for a beneficial SPADE exposure.} Under the fixed-response Poisson assumptions above, with $F_{\nu\nu}^{(\DI)}$ positive definite and $I_{\DI}>0$, there is an allocation $0<t<1$ whose effective depth information exceeds that of DI alone if and only if

\begin{equation}\label{eq:derivative}
\eta v_{\DI}\tr F^{(\mathrm{SPADE})}v_{\DI}>I_{\DI}.
\end{equation}
This specializes local $c$-optimal design: loss-adjusted SPADE information along DI's hardest direction must exceed information displaced from DI. Effective information $I_\eff[F(t)]$ is concave \cite{pukelsheim}; a positive initial slope gives improvement and a nonpositive slope rules it out for every allocation. Appendix~\ref{app:allocation} proves this. If $v_{\DI}\tr F^{(\mathrm{SPADE})}v_{\DI}>0$, the transmission threshold is
\begin{equation*}
\eta>\frac{I_{\DI}}{v_{\DI}\tr F^{(\mathrm{SPADE})}v_{\DI}}.
\end{equation*}
The denominator is lossless-SPADE information along the same DI direction, not SPADE's independently minimized effective information. The inequality is strict.

With known nuisance parameters this reduces to $\eta F_{\alpha\alpha}^{(\mathrm{SPADE})}>F_{\alpha\alpha}^{(\DI)}$. With unknowns, SPADE can also resolve their coupling. The criterion establishes existence of a beneficial allocation, not its optimum or improvement at every $t>0$.

Section~\ref{sec:allocation_results} tests fixed-response predictions. Complete acquisition additionally includes a data-dependent center, localization error, readout/alignment/calibration time, and calibration parameters potentially unidentifiable from DI alone. These fall outside (\ref{eq:allocation}); we evaluate achieved depth RMSE, including bias, by Monte Carlo.

\section{Numerical Experiments}\label{sec:experiments}\label{sec:adaptive}

\begin{figure*}[!t]\centering
\includegraphics[width=\textwidth]{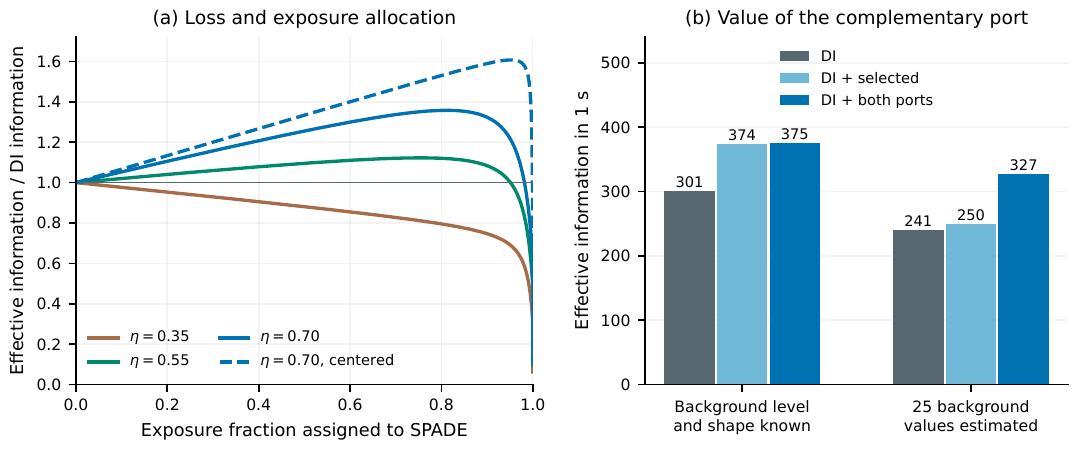}
\caption{Fixed-response design with an estimated C4 background. (a) Effective depth information relative to DI versus mode exposure fraction. Solid curves use the nominal center at labeled transmissions; the blue dashed curve uses the true center at transmission 0.70. The horizontal line denotes equal information. (b) Information in 1 s for DI and 20/80 DI--mode acquisition, retaining one or both ports. Background level/shape are fixed in the known group; all 25 values are fitted in the estimated group. The complement primarily constrains background. Both panels include 12 nm response and known 1\% crosstalk, with unknown target brightness, slope, and position. Responses/calibration are fixed; Fig.~\ref{fig:depth_confirmation} tests complete acquisition.}\label{fig:allocation_validation}
\end{figure*}

We test optical selectivity (Fig.~\ref{fig:retention}), allocation and complementary-port information (Fig.~\ref{fig:allocation_validation}), complete adaptive estimation (Fig.~\ref{fig:depth_confirmation}), and material/geometric dependence (Fig.~\ref{fig:external}). Fixed-response Fisher calculations isolate design mechanisms; Monte Carlo RMSE includes bias, random localization, and operating costs without equating achieved error to an information bound.

\subsection{Experimental setup and evaluation protocol}\label{sec:setup}
Pure DI observes throughout its allocated scene time. Adaptive DI estimates a center, then records both SPADE-port spectra. Both estimate the same $\alpha$. Each fixed-scene trial regenerates photon/dark counts, read noise, and calibration; adaptive initial data also regenerate the command. The result is repeated-acquisition precision for measured spectra and synthetic geometry.

C4 uses measured calcite co2004/dolomite cod2005 reflectances (Fig.~\ref{fig:overview}). Interpolation gives 25 samples; the target's upper convex envelope supplies a piecewise-linear continuum across the trough, followed by (\ref{eq:profile_definition}). Reference depth 0.25053 is simulation truth except in depth scans, and is withheld from fitting.

\begin{figure*}[t]\centering
\includegraphics[width=\textwidth]{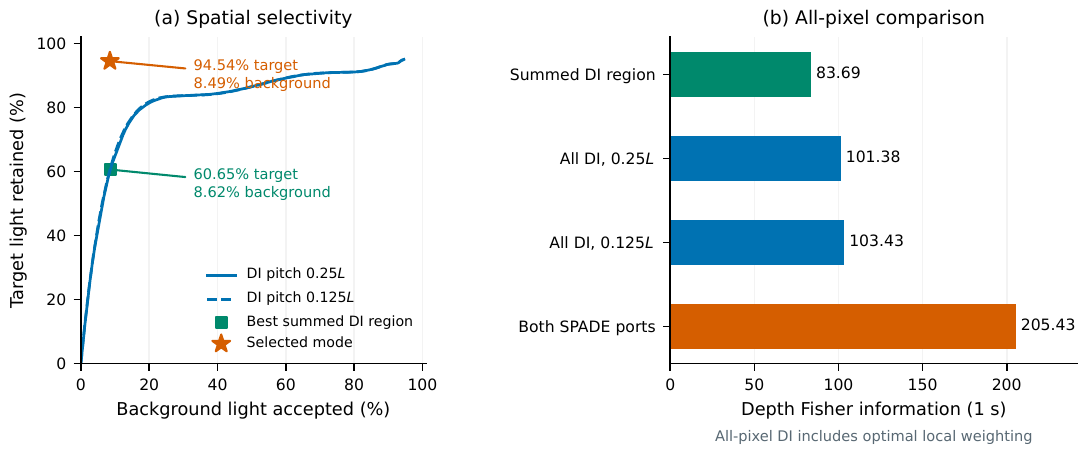}
\caption{Ideal C4 comparison at 2335.97 nm, with $L=1.09975$ mm. (a) Target retention versus background acceptance for circular DI regions at pitches $0.25L$ and $0.125L$. The marked region maximizes summed-count Fisher information. (b) Information for equal 1 s exposures: the best summed region, all DI pixels/optimal local weights, and both mode ports. The mode center is $(0.5L,0)$ and target center $(0.62L,0.08L)$. Brightness, position, and background are known; extra transmission is unity and crosstalk zero. SPADE exceeds even the finer-DI information.}\label{fig:retention}
\end{figure*}

Table~\ref{tab:status} separates supplied, inferred, and calibrated quantities. Reference spectra and prior imaging/sample description provide target shape, continuum, and geometry before the 10 s budget; their preparation is assumed, not simulated. Initial DI commands the mode center, then the final fit combines those same data, both ports, and calibration [Fig.~\ref{fig:overview}(c)].

\begin{table}[t]\centering\footnotesize
\caption{Parameter status in the main C4 acquisition (Fig.~\ref{fig:depth_confirmation}).}\label{tab:status}
\begin{tabular}{@{}>{\raggedright\arraybackslash}p{0.29\columnwidth}p{0.64\columnwidth}@{}}\toprule
Status & Quantities \\\midrule
Supplied & Target continuum $C_\ell$, normalized feature $h_\ell$, size and shape; background spatial weight; optical and spectral responses \\
Estimated from scene & Depth $\alpha$, target brightness, continuum slope, two position coordinates; 25 positive background spectral values \\
Jointly calibrated & Sorter transmission $\eta$, port leakage $\epsilon$, and two calibration-source intensities \\
Fixed assumptions & Target curvature $q_T=0$; one background spectrum shared over space; exact execution of the estimated alignment command \\
Separate material test & Fig.~\ref{fig:external} supplies both spectral shapes and fits their brightnesses; target slope and curvature are fixed \\
\bottomrule\end{tabular}\end{table}

Table~\ref{tab:parameters} gives C4 settings. With $L=\lambda_{\rm ref}z/D=1.09975$ mm, the target is approximately $0.275\times0.137$ mm. Diffraction varies with wavelength at fixed $L$. Background weight $1+0.15u_x$ excludes the target; the larger DI field collects diffracted tails beyond the source square.

\begin{table}[!htbp]\centering\small
\caption{Nominal C4 scene and instrument settings. Lengths are in the object plane; $L=1.09975$ mm. Receiver-specific operating times are in Table~\ref{tab:time}.}\label{tab:parameters}
\begin{tabular}{@{}p{0.43\columnwidth}p{0.50\columnwidth}@{}}\toprule
Quantity & Setting \\\midrule
Range; aperture diameter & 10 m; 2 cm \\
Reference wavelength & 2199.5 nm \\
Target ellipse, full axes & $0.25L\times0.125L$ \\
True target center & $(0.62L,0.08L)$ \\
Nominal mode center & $(0.5L,0)$ \\
Source field; DI field & $[-2L,2L]^2$; $[-4L,4L]^2$ \\
Spectral grid & 25 channels, 2249.18--2397.84 nm \\
Spectral-response FWHM & 12 nm \\
Incident irradiance & 10 W/m$^2$, flat over 2100--2400 nm \\
Common optical/detector efficiency & 0.2 \\
DI spatial sampling & $16\times16$ pixels, pitch $0.5L$ \\
Detector pixels per mode port and wavelength & 1 selected; 16 complement \\
Mode transmission; crosstalk & 0.7; 0.01 \\
Dark current & 0.05 electrons/s/pixel \\
Read noise & 18 electrons per CDS frame \\
Readout outputs; per-output rate & 32; $10^5$ pixels/s \\
\bottomrule\end{tabular}\end{table}

Counts follow (\ref{eq:di})--(\ref{eq:mean_main}), then the 12 nm full width at half maximum (FWHM) response and port mixing. Shared surface, illumination, efficiency, and validated radiometric/noise implementation are specified in Appendices~\ref{app:optics}--\ref{app:validation}; hardware performance remains untested.

Both complete acquisitions use Table~\ref{tab:parameters}'s DI sampling; sixteen complement pixels are summed per wavelength into one spectrum. Figure~\ref{fig:retention}'s finer-DI comparison isolates optical sampling, without optimizing the full instrument's pixel/readout architecture.

Following Grace \emph{et al.} \cite{grace}, initial DI jointly fits depth, target brightness/slope, position, and 25 background values in the main C4 test, with the same scene freedom for pure DI. Material tests instead supply both shapes and fit depth, brightnesses, and position. Only the estimated center commands SPADE, assumed accurately executed; alignment error comes from noisy localization. Truth generates data but is never supplied to the fit/command.

Joint fitting combines initial DI, mode, and calibration counts at the recorded command. True position remains unknown. Initial DI contributes once, without an independent position prior (Appendix~\ref{app:noise}).

Table~\ref{tab:time} charges 10 s to either receiver: adaptive science receives 8.8 s versus 9.5 s for DI, before segment-wise readout deductions. Shared reference preparation precedes this budget; schedules remain fixed across main material tests.

Both final fits bound depth to $[0,0.99]$ and each position coordinate to $[-0.6L,0.6L]$ relative to $(0.5L,0)$. Adaptive fitting additionally estimates calibration parameters, accounting for their uncertainty and shorter exposure.

\begin{table}[!htbp]\centering
\caption{Observation-time budget in seconds.}\label{tab:time}
\begin{tabular}{lrr}\toprule Segment & DI & Adaptive \\\midrule
DI scene measurement & 9.5 & 1.8 \\
Processing and alignment & 0 & 0.2 \\
SPADE scene measurement & 0 & 7.0 \\
Dark reference & 0.5 & 0.5 \\
Mode calibration & 0 & 0.5 \\\midrule Total & 10.0 & 10.0 \\\bottomrule
\end{tabular}\end{table}

True mode transmission/crosstalk are 0.7/0.01, the latter the probability of switching ports after transmission. Two ideal orthogonal calibration inputs are each observed by a monitor and both ports. Their finite noisy counts jointly estimate transmission, leakage, and unknown input intensities. Appendix~\ref{app:noise} gives all six means, monitor/input assumptions, and estimation errors; Table~\ref{tab:stages} reports code-extracted frame accounting.

Poisson scene/dark counts and Gaussian read noise generate observations. Fits use the matching Gaussian mean/variance likelihood (\ref{eq:fit}); several starts are compared and the lowest objective retained (Appendix~\ref{app:noise}).

We evaluate absorption estimates using root-mean-square error and its ratio between receivers:
\begin{equation}\label{eq:rmse_metric}
\begin{aligned}
\RMSE&=\left[\frac{1}{N_{\rm trial}}\sum_{n=1}^{N_{\rm trial}}
(\widehat\alpha_n-\alpha_0)^2\right]^{1/2},\\
\text{RMSE ratio}&=\frac{\RMSE_{\rm adaptive}}{\RMSE_{\DI}}.
\end{aligned}
\end{equation}
Here $N_{\rm trial}$ counts acquisitions, $\alpha_0$ is truth, and $\widehat\alpha_n$ its trial estimate. RMSE combines bias/random error and retains every estimate, including bounds. Ratios below one favor adaptive sensing but must be assessed alongside absolute errors.

Bootstrap resamples saved estimates and recomputes ratios. An interval wholly below one supports adaptive improvement; overlap with one leaves ordering unresolved. Fixed-reference resampling conditions on those references, whereas reference-cluster resampling also captures the specified perturbation variation. Neither includes unmodeled field/instrument uncertainty (Appendix~\ref{app:validation}).

\subsection{Ideal optical selectivity and absorption information}\label{sec:ideal_results}
We isolate optical selection from localization/calibration/fitting by estimating only depth: a 1 s exposure at 2335.97 nm (sampled band maximum), known brightness/position/background, unity additional transmission, zero crosstalk, and photon noise only. Common radiometric efficiency remains.

Forward-model target/background rates give their retained fractions. DI sums pixels whose centers fall inside increasing circular radii about the true target; SPADE uses fixed center $(0.5L,0)$. Fractions refer to each material's total light before spatial selection, not its proportion of the recorded mixture. The detector observes only summed contributions.

Figure~\ref{fig:retention}(a) shows 94.54\% target retention at 8.49\% background acceptance for the selected mode, versus 60.65\%/8.62\% for the best summed DI circle. At nearly equal background, larger retained target response increases information as in (\ref{eq:retention_fisher}).

To include the strongest DI baseline, we compare the best summed region, all pixels, and both ports. A region contributes squared derivative of its summed mean divided by that mean; full data sum individual pixel/port information. The all-pixel value also equals optimal local weighting (\ref{eq:weights}). These are deterministic information calculations, not fitted-trial errors.

Figure~\ref{fig:retention}(b) gives information 83.69 for the best summed DI circle, 101.38 for all pixels/optimal weighting, and 205.43 for both SPADE ports. Halving DI pitch from $0.25L$ to $0.125L$ raises it to 103.43; SPADE retains approximately 1.99 times this finer-DI information. We next add unknowns and acquisition costs.

Doubling DI field halfwidth from $4L$ to $8L$ at pitch $0.25L$ raises information only from 101.38 to 101.97 (0.58\%), without renormalizing lost light. Finite coverage therefore explains little of this single-band difference. Exposure is fixed and detector noise excluded; extra pixels are not cost-free in a full instrument.

\begin{figure*}[t]\centering
\includegraphics[width=\textwidth]{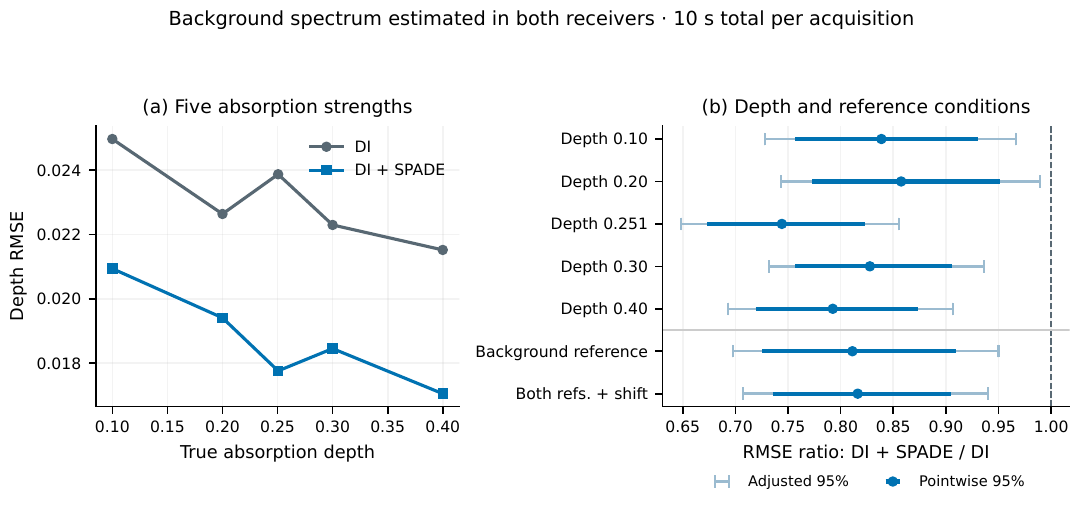}
\caption{Complete 10 s acquisitions with the same freely estimated background in both receivers, 400 trials per arm. (a) Depth RMSE at five true depths with fixed target continuum/profile; lines guide the eye. (b) Adaptive/DI ratios with pointwise 95\% intervals (dark) and seven-comparison Bonferroni intervals (light); values below one favor adaptive acquisition. Reference-error tests use new perturbations of one standard error of the mean (SEM), with 12 nm correlation; the last also applies a shared offset within $\pm0.5$ nm. Adaptive trials include DI localization, alignment time, calibration, and detector costs.}\label{fig:depth_confirmation}
\end{figure*}

\subsection{Time allocation and complementary-port information}\label{sec:allocation_results}
We test (\ref{eq:derivative}) in C4 with unknown depth, target brightness/slope, position, and 25 background values. The 12 nm response, known 1\% crosstalk, and known transmission enter fixed-response independent Poisson means, isolating design from noisy localization/costs.

Figure~\ref{fig:allocation_validation}(a) confirms nominal-center transmission threshold 0.4559. At 0.70, optimal mode fraction is 0.8115 and information rises 35.8\%; target-centered modes lower the threshold to 0.4192. These local optima do not optimize Table~\ref{tab:time}'s separately fixed adaptive schedule, which pays localization, alignment, calibration, and readout. Independent derivative/projection checks verify the criterion (Appendix~\ref{app:validation}).

For a 20/80 acquisition with known background level/shape, adding the complement changes information only from 373.8 to 375.5 [Fig.~\ref{fig:allocation_validation}(b)]. With unknown background it rises from 250.2 to 327.1, versus DI's 240.9. The complement constrains background while the selected port retains target response. Selected-port-only SPADE without DI has zero effective depth information: free background can reproduce absorption changes. Complementary spatial mixtures remove this ambiguity.

\begin{figure*}[!t]\centering
\includegraphics[width=\textwidth]{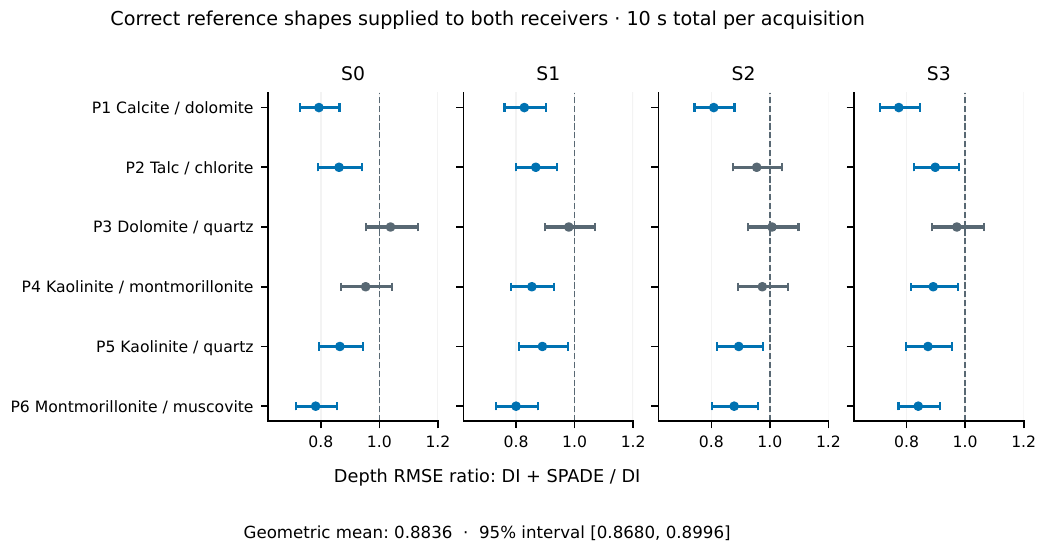}
\caption{Ten additional specimens with fixed receiver settings. Six pairs are each tested in four geometries, with 500 trials per receiver and 10 s budgets. Bars are pointwise 95\% bootstrap intervals: blue markers indicate intervals below one, gray intervals containing one. The footer gives the equal-weight geometric mean of all 24 ratios and its interval. Correct component spectra are supplied in every condition, testing material/geometric dependence separately from unknown-background confirmation.}\label{fig:external}
\end{figure*}

\subsection{Complete acquisition with an estimated background}\label{sec:depth_confirmation}\label{sec:joint_results}

The independent C4 confirmation includes localization, calibration, detector noise, and background learning with equal scene freedom and 10 s budgets. Before generation, the 25-background model, bounds, and optimizer were frozen. Five true depths are 0.10, 0.20, 0.25053, 0.30, and 0.40; only $\alpha$ changes at fixed $C_\ell,h_\ell$, without flux renormalization. Every fit starts from $\alpha=0.3$.

Each condition has 400 acquisitions per receiver. At original depth, two more tests use new background-only 1 SEM errors or both-reference 1 SEM errors plus shared offset within $\pm0.5$ nm, each with 40 reference pairs and ten noise repeats. Exact references use independent-arm trial bootstrap; perturbed references use paired reference-cluster bootstrap, all with 500\,000 resamples (Appendix~\ref{app:references}). Background-only replacement tests coordinate/optimization stability in a fully parameterized spectrum, not unmodeled physical background (Appendix~\ref{app:reference_invariance}).

Adaptive RMSE is lower at every depth (Fig.~\ref{fig:depth_confirmation}): ratios 0.744--0.858 give 14\%--26\% reduction, and all seven adjusted intervals are below one. Initial-position 95th-percentile error falls from $0.391L$ at depth 0.10 to $0.111L$ at 0.40, so the shallowest band also challenges localization. All final depths stay interior; initial/boundary trials remain included (Appendix~\ref{app:independent_depth}).

\subsection{Additional mineral specimens and target geometries}
A separate supplied-spectrum test isolates material/geometric dependence from background learning. Ten USGS specimens absent from design form six pairs (Table~\ref{tab:pairs}), specified before receiver evaluation and all passing coverage/depth checks. Fits estimate depth, two brightnesses, and position under unchanged timing, pitch, mode, and transmission.

Each pair uses four target/brightness geometries: C4 ellipse/gradient S0, disk/uniform S1, three-lobed/quadratic-background S2, and larger rotated ellipse S3. Initial offsets vary too. Every one of 24 conditions uses 500 independent trials per receiver, repeating adaptive localization and fitting (Appendix~\ref{app:validation}).

Pairs include carbonates, talc/chlorite, and clays; some specimens recur. P1--P3 use 25 samples over 2249.18--2397.84 nm; P4--P6 over 2099.95--2249.18 nm. Correct references/geometry are supplied while depth, brightness, and position remain unknown. This USGS-library benchmark is separate from unknown-background C4 confirmation.

\begin{table}[!htbp]\centering\small
\caption{Additional target/background pairs. Specimen identifiers are from the USGS library.}\label{tab:pairs}
\begin{tabular}{lll}\toprule Pair & Target & Background \\\midrule
P1 & Calcite GDS304 & Dolomite ML97-3 \\
P2 & Talc WS659 & Chlorite GDS307 \\
P3 & Dolomite ML97-3 & Quartz HS32 \\
P4 & Kaolinite CM9 & Montmorillonite SAz-1 \\
P5 & Kaolinite KGa-1 & Quartz HS32 \\
P6 & Montmorillonite SCa-2 & Muscovite GDS113 \\\bottomrule
\end{tabular}\end{table}

Figure~\ref{fig:external} gives condition-wise ratios and pointwise 95\% intervals. Exponentiating the mean log ratio gives an equal-weight geometric mean over this fixed benchmark, not pooled depth RMSE or arbitrary-scene population performance.

The geometric mean is 0.8836, interval $[0.8680,0.8996]$. Seventeen pointwise intervals and eleven Bonferroni-adjusted intervals lie below one, stable across bootstrap batches (Appendix~\ref{app:validation}). Table~\ref{tab:material_absolute} gives absolute errors; index S1 in \nolinkurl{SUPPLEMENT_INDEX.html} inside \nolinkurl{manuscript_experiments.zip} locates \nolinkurl{all_24_conditions.csv}, including adjusted intervals and boundary fractions. No condition's interval is wholly above one; overlap leaves ordering unresolved.

Calcite/dolomite, kaolinite/quartz, and montmorillonite/muscovite improve in all geometries pointwise; dolomite/quartz is nearly neutral throughout. An absorption feature alone does not ensure gain: spectra and geometry matter.

\subsection{Target continuum and spatial background}\label{sec:scope}
The main test supplies target shape/geometry, fits its continuum slope, and fixes curvature to zero. Table~\ref{tab:sensitivity} shows that adding unknown curvature reduces local DI/mixed information from 240.9/327.1 to 28.9/40.7, enlarging inverse-square-root SD bounds 2.89/2.83 times. Depth and curvature responses have partial absolute correlation about 0.94: continuum bending can mimic absorption. These local Poisson bounds reveal absolute-precision dependence on continuum knowledge; they are not complete-acquisition RMSE.

\begin{table}[t]\centering\footnotesize
\caption{Target-continuum sensitivity in C4. Information and RMSE are separate diagnostics.}\label{tab:sensitivity}
\begin{tabular}{@{}lrr@{}}\toprule
Local information in 1 s & DI & Mixed \\\midrule
Curvature fixed & 240.9 & 327.1 \\
Curvature also estimated & 28.9 & 40.7 \\\midrule
Complete acquisition (400/arm) & DI & Adaptive \\\midrule
Curvature omitted: bias & 0.00637 & 0.00385 \\
\quad RMSE & 0.02283 & 0.01978 \\
\multicolumn{3}{@{}l}{Curvature fitted within $\pm0.03$} \\
\quad Bias & 0.00375 & 0.00151 \\
\quad RMSE & 0.02947 & 0.02685 \\\bottomrule
\end{tabular}
\par\smallskip\raggedright
The continuum coefficients are defined in (\ref{eq:continuum_extension}). Both acquisition rows generate target/background slopes $+0.01/-0.01$ and curvatures $+0.01/-0.01$. The background remains freely estimated. Ratios are 0.867 and 0.911. The fitted target-curvature bounds $q_T\in[-0.03,0.03]$ are an illustrative diagnostic setting, not an independently calibrated uncertainty range.
\end{table}

Two exploratory C4 acquisitions retain the 10 s schedules and small true continuum curvature. Omitting target curvature gives adaptive/DI RMSE 0.01978/0.02283; fitting it lowers bias but raises RMSE to 0.02685/0.02947. The example numerical bound $q_T\in[-0.03,0.03]$ has no independent-reference basis. Curvature touches bounds in 263/400 adaptive and 305/400 DI fits: these bounded-estimator diagnostics do not establish unconstrained recovery. Their records are separate from Fig.~\ref{fig:depth_confirmation}.

A free shared spectrum does not model spatial composition gradients. Appendix~\ref{app:spatial_background} mixes quartz into dolomite. At maximum fraction 0.02, DI/adaptive biases are $-0.03035/-0.02291$; adaptive RMSE is 0.02943 versus uniform-control 0.01866. At maximum 0.05 it reverses ordering: DI/adaptive RMSE 0.05202/0.09780, ratio 1.880. Fractions denote mixture composition, not reflectance-error percentages. Stronger cases show optimization ambiguity and position bounds at 0.10/0.20 (Table~\ref{tab:spatial_background}); ratios describe this mismatched inference pipeline, not fundamental optical limits.

Application requires a representative target reference resolving the band and both shoulders under consistent calibration \cite{koerting}, plus target geometry from prior imaging/sample characterization. Absolute RMSE conditions on supplied feature/continuum shape with fitted slope and fixed curvature. Diagnostics establish sensitivity without a numerical reference-accuracy specification.

Prior composition information and initial DI residuals can flag shared-background mismatch. Spatially structured spectral residuals, position bounds, or start sensitivity prompt a background check; these are qualitative warnings, not a validated threshold/test. C4's $0.275\times0.137$ mm target at 10 m and 2 cm aperture gives a scale for laboratory reference and broadband-sorter tests. Within supplied-target assumptions, the design supports local feature-strength comparison without a classification/abundance model.

\section{Conclusions}\label{sec:conclusion}
We developed a quantum-inspired receiver framework for the absorption depth of a selected mineral patch in overlapping light. Recovering the target's own feature strength, rather than mixed-pixel contrast, supports quantitative comparisons when the mineralogical interpretation is established.

Finite reflecting regions and structured spectra connect mode selection to mineral spectroscopy. Retaining target response while rejecting background shot noise can improve on all-pixel DI, including optimal local weighting. Joint inference also requires depth to be distinguishable from brightness/position. Specializing local optimal-design theory yields a necessary and sufficient beneficial-exposure condition under fixed-response Poisson measurements, a loss threshold, and an explanation of complementary DI--SPADE information.

Complete 10 s acquisitions include localization uncertainty, alignment time, calibration, and detector noise. With a fitted spatially shared background, independent five-depth simulations reduce RMSE by 14\%--26\%. Target shape/geometry are supplied, slope fitted, and curvature fixed. Controlled reference tests examine coordinate stability and small target-template errors. The separate supplied-spectrum 24-condition benchmark gives equal-weight geometric-mean ratio 0.884, establishing material/geometric dependence without conflating it with unknown-background confirmation.

Initial DI and the complementary port constrain the unknown shared spectrum while the selected port preserves absorption response. The analysis connects background characterization and optical rejection in one design. Continuum and spatial-composition diagnostics identify what must be characterized separately for a new scene.

For proximal sensing, this strategy designs the optical measurement around a selected feature's precision: mode selection controls accompanying photon noise, and joint inference separates the feature from background. The allocation criterion and complete-acquisition model support laboratory receiver choices and connect quantum-inspired measurement to quantitative imaging spectroscopy.

\section*{Data availability}
The simulations use public USGS reflectance spectra \cite{usgs}. Code and simulation results are provided in the supplementary archive. 

\section*{Funding Declaration}
This work was supported by the National Natural Science Foundation of China under Grant 62571475.

\appendices
\section{Optical response and radiometric normalization}\label{app:optics}
Multiplying the dimensionless collection areas $A_c$ and $A_{c\ell j}$ by $L^2$ gives physical weighted areas. For a clear circular pupil, the normalized Airy amplitude PSF at radial distance $r=|\mathbf x-\mathbf u|$ (in units of $L$) is
\begin{equation}\label{eq:airy}
\psi_{\ell,\mathbf u}(\mathbf x)=
\frac{\sqrt{\pi}\lambda_{\rm ref}}{2\lambda_\ell}
\frac{2J_1(\pi r\lambda_{\rm ref}/\lambda_\ell)}{\pi r\lambda_{\rm ref}/\lambda_\ell},
\end{equation}
Here $J_1$ is the first-order Bessel function. The factor $\lambda_{\rm ref}/\lambda_\ell$ changes diffraction width at fixed $L$, while the prefactor normalizes integrated intensity to one. The Bessel ratio tends to one at the origin. Propagation preserves field inner products, and the overlap of two displaced PSFs equals the normalized Fourier transform of the clear pupil. With source-to-mode displacement $d=|\mathbf u-\mathbf u_0|$, this gives
\begin{equation}\label{eq:airy_overlap}
p_0(\mathbf u,\ell)=
\left[\frac{2J_1(\pi d\lambda_{\rm ref}/\lambda_\ell)}{\pi d\lambda_{\rm ref}/\lambda_\ell}\right]^2.
\end{equation}
Unlike the PSF radius $r$, $d$ measures source displacement from the mode center. The bracket is a normalized field overlap; its square gives $p_0=1$ at $d=0$. We integrate this probability over every target and background surface point, retaining finite target geometry.

Lambertian radiance is $E_\ell R_{c\ell}/\pi$, where $E_\ell$ is spectral irradiance. Multiplying by physical area $L^2\dd^2\mathbf u$, aperture solid angle $\pi D^2/(4z^2)$, and bin width $\Delta\lambda_\ell$, then dividing by photon energy $h_{\rm P}c_0/\lambda_\ell$, gives the rate. Including common detection efficiency $\eta_0$ yields
\begin{equation}\label{eq:rate}
K_\ell=\eta_0\frac{D^2}{4z^2}L^2
E_\ell\Delta\lambda_\ell\frac{\lambda_\ell}{h_{\rm P}c_0},
\end{equation}
Here $h_{\rm P}$ and $c_0$ are Planck's constant and the speed of light. The aperture and Lambertian factors combine as $D^2/(4z^2)$. Use consistent spectral units and meters for wavelength in the photon-energy factor. The common efficiency $\eta_0$ differs from the additional mode transmission $\eta$. Receiver losses are applied separately; unrecorded DI tails are not renormalized.

Mode-crosstalk analyses motivate retaining nonideal port mixing \cite{gessner2020}. The spectral response has Gaussian FWHM 12 nm and a prefiltered input range. Let $H_{k\ell}$ be the fraction of input-bin $\ell$ photons recorded in output wavelength channel $k$, obtained by integrating that Gaussian over the channel. Edge loss permits $\sum_k H_{k\ell}<1$ and is not renormalized. Symmetric port-switching probability $\epsilon=0.01$ and additional transmission $\eta=0.7$ are calibrated. For unfiltered mode means $\mu_{\ell j}$ from (\ref{eq:mean_main}) at the live exposure, the final optical means are
\begin{equation}\label{eq:instrument_response}
\begin{aligned}
\mu^{\rm det}_{k0}&=\eta\sum_\ell H_{k\ell}
\left[(1-\epsilon)\mu_{\ell0}+\epsilon\mu_{\ell\perp}\right],\\
\mu^{\rm det}_{k\perp}&=\eta\sum_\ell H_{k\ell}
\left[\epsilon\mu_{\ell0}+(1-\epsilon)\mu_{\ell\perp}\right].
\end{aligned}
\end{equation}
Each output combines its retained light with leakage from the other port before summing over input wavelengths; $\eta$ reduces both means. These means exclude detector noise. DI uses the same $H_{k\ell}$ without port mixing or additional mode loss.

Table~\ref{tab:parameters}'s range/aperture and flat 2100--2400 nm illumination give $L=1.09975$ mm and target rate approximately $2.48\times10^5$ electrons/s before spatial selection. Irradiance is an assumed active broadband source, not a measured lamp or solar spectrum.

Programmable mode conversion and sorting provide optical implementations of field decomposition \cite{morizur2010,fontaine2019}. The geometric etendue (collection area times solid angle) is approximately $6.08\times10^{-11}$ m$^2$ sr. Sixteen 18 $\mu$m pixels with an assumed F/2 relay have sufficient complement-port capacity; this does not establish low-loss broadband sorting. DI requires simultaneous spatial/spectral collection, e.g., an integral-field arrangement, whose optical losses belong in its transmission.

\section{Detector model and numerical inference}\label{app:noise}
The detector uses dark current 0.05 electrons/s/pixel, 18-electron read noise per correlated double sampling (CDS) frame, and 80\,000-electron full well, with frames scheduled below 70\% capacity. Selected and complement ports occupy one and sixteen pixels per wavelength. H2RG noise specifications \cite{h2rg} and a commercial shortwave-infrared (SWIR) spectrometer bandwidth \cite{specim} provide references. The assumed 32 parallel electronic outputs are distinct from optical ports.

For $N_{\rm pix}$ physical pixels (including spectral channels), $n_{\rm out}$ parallel electronic outputs at $10^5$ pixels/s each, and two read passes per CDS frame, assumed frame dead time is
\begin{equation}\label{eq:readout_time}
t_{\rm dead}=\frac{2N_{\rm pix}}{n_{\rm out}(10^5\ {\rm s}^{-1})}+10^{-4}\ {\rm s}.
\end{equation}
The last term is a 0.1 ms frame overhead; live integration subtracts all frame dead times from the segment duration. Dark and calibration segments have their own schedules. Two passes form one CDS frame, so summing $n$ frames over $k$ pixels contributes read variance $nk(18\ {\rm e^-})^2$, without a second factor of two \cite{h2rg}. This assumed sequence couples readout configuration to both exposure and noise.

Schedules are fixed before acquisition. Conservative rate bounds are $1.5K_{\max}p^2$ for DI and $1.5K_{\max}\max[4(\lambda_{\max}/\lambda_{\rm ref})^2/\pi,1]$ for mode detection, with $K_{\max}=\max_\ell K_\ell$, normalized pitch $p=0.5$, and calibration bound $3\times10^7$ electrons/s. For segment duration $T_s$, rate bound $r_{\max}$, dark rate $d$, and full well $W$,
\begin{align}
n&=\max\!\left\{1,\left\lceil\frac{T_s}{0.7W/(r_{\max}+d)+t_{\rm dead}}\right\rceil\right\},\\
T_{\rm live}&=T_s-nt_{\rm dead}.
\end{align}
The initial position estimate does not change this schedule. Table~\ref{tab:stages} reports C4 accounting. Aggregate output-rate access is assumed; detector window packing and controller timing are not measured.

\begin{table*}[t]\centering\footnotesize
\caption{C4 acquisition accounting from the code. Read RMS is accumulated per physical pixel before dark subtraction; multiply it by four for a sum of sixteen complement pixels. Calibration and its dark row each occur twice. The 0.2 s alignment interval has no exposure or reads.}\label{tab:stages}
\begin{tabular}{lrrrrrrr}\toprule
Segment & Physical & Wall time & CDS & Live time & Total dead & Read RMS & Dark scale \\
& pixels & (s) & frames & (s) & time (s) & (electrons) & $\rho$ \\\midrule
DI scene & 6400 & 9.500 & 38 & 9.344200 & 0.155800 & 110.96 & 18.8429 \\
DI dark & 6400 & 0.500 & 1 & 0.495900 & 0.004100 & 18.00 & --- \\
Adaptive initial DI & 6400 & 1.800 & 8 & 1.767200 & 0.032800 & 50.91 & 3.5655 \\
Adaptive mode ports & 425 & 7.000 & 170 & 6.937844 & 0.062156 & 234.69 & 13.9979 \\
Adaptive science dark & 6825 & 0.500 & 1 & 0.495634 & 0.004366 & 18.00 & --- \\
Calibration, per input & 18 & 0.125 & 64 & 0.117880 & 0.007120 & 144.00 & 0.9439 \\
Calibration dark, per input & 18 & 0.125 & 1 & 0.124889 & 0.000111 & 18.00 & --- \\
\bottomrule\end{tabular}\end{table*}

Scene and dark electrons are independent Poisson counts; read noise is Gaussian. The dark-subtracted observation is
\begin{equation}\label{eq:noise}
Y=N_s+N_d+e_r-\rho(N_{d,\rm ref}+e_{r,\rm ref}),
\end{equation}
Here $N_s,N_d$ are scene and dark counts; $e_r$ is zero-mean accumulated read noise. Subscript $\mathrm{ref}$ denotes an independent dark exposure, scaled by the scene/reference live-time ratio $\rho$. Spatial/wavelength indices are suppressed. $Y$ is in electrons and may be negative.

Let $\mu_s,\mu_d,\mu_{d,\rm ref}$ denote mean counts and $\sigma_r^2,\sigma_{r,\rm ref}^2$ accumulated read variances. Matching dark rates gives $\mu_d=\rho\mu_{d,\rm ref}$, hence $\mathbb E[Y]=\mu_s$ and
\begin{equation}\label{eq:noise_variance}
\begin{aligned}
V=\operatorname{Var}(Y)
&=\mu_s+\mu_d+\sigma_r^2\\
&\quad+\rho^2(\mu_{d,\rm ref}+\sigma_{r,\rm ref}^2).
\end{aligned}
\end{equation}
Independent reference subtraction adds the last variance term, including squared scaling $\rho^2$. The adaptive dark slot simultaneously reads disjoint groups of 6400 DI and $25(1+16)=425$ mode pixels. Each channel subtracts its own dark realization once from accumulated science counts. Each calibration input has a separate dark exposure. Thus no dark value is shared across DI, mode, and calibration observations, and no corresponding cross-observation covariance is omitted conditional on the command and parameters. Correlated electronics noise is outside this independent-pixel model.

Calibration uses ideal pure inputs in the selected mode and its orthogonal complement. With unknown intensity multipliers $a_0,a_1$ (true values one), the expected dark-subtracted counts in the monitor, selected port, and complement are
\begin{align}\label{eq:calibration_means}
\boldsymbol\mu_{{\rm cal},0}&=a_0N_{\rm cal}
  \begin{pmatrix}1\\\eta(1-\epsilon)\\\eta\epsilon\end{pmatrix},\\
\boldsymbol\mu_{{\rm cal},1}&=a_1N_{\rm cal}
  \begin{pmatrix}1\\\eta\epsilon\\\eta(1-\epsilon)\end{pmatrix}.
\end{align}
At unit intensity, the normalized pre-sorter monitor mean is $N_{\rm cal}=10^7(0.11788)=1\,178\,800$ electrons, with known monitor-to-port gain ratio. Monitor and selected channels use one pixel each, the complement sixteen, with Table~\ref{tab:stages}'s dark/CDS model. At $\eta=0.7$, $\epsilon=0.01$, the mean vectors are $(1\,178\,800,816\,908.4,8251.6)$ and $(1\,178\,800,8251.6,816\,908.4)$.

The six counts constrain $\eta,\epsilon,a_0,a_1$, giving calibration-only local standard-deviation bounds $7.64\times10^{-4}$ and $1.97\times10^{-4}$ for $\eta,\epsilon$. Nominal-depth joint scene/calibration estimates have respective biases $-4.7\times10^{-6}$, $5.7\times10^{-6}$ and RMSEs $5.2\times10^{-5}$, $3.6\times10^{-5}$. These smaller errors also use scene information; per-condition statistics are archived.

A single wavelength-independent $(\eta,\epsilon)$ pair and ideal pure inputs are assumed. Calibration wavelength distribution, input-preparation error, monitor gain-ratio uncertainty, and hardware switching are not specified by measured hardware.

The noisy dark reference contributes median variance fractions of 67.5\% in DI and 2.1\%/3.3\% in the mode ports. A local diagnostic fixes all schedules, read costs, calibration noise, and the nominal mode center, replacing only science-dark reference uncertainty by a known mean; science dark shot noise remains. Gaussian mean-and-variance information changes from 1943.6 to 2214.6 for DI and 2723.6 to 2786.6 for the mixed acquisition, reducing their ratio from 1.401 to 1.258. Thus noisy dark subtraction affects the gain size but does not create it in this diagnostic. This is neither Monte Carlo RMSE nor a reoptimized schedule.

For scene and calibration observations $Y_i$, let $\mu_i(\boldsymbol\theta)$ include the final optical response (\ref{eq:instrument_response}) and let $V_i(\boldsymbol\theta)$ be (\ref{eq:noise_variance}). A moment-matched Gaussian approximation gives the negative log-likelihood, up to parameter-independent constants,
\begin{equation}\label{eq:fit}
\mathcal L(\boldsymbol\theta)=\frac12\sum_i\left\{
\frac{[Y_i-\mu_i(\boldsymbol\theta)]^2}{V_i(\boldsymbol\theta)}
+\log V_i(\boldsymbol\theta)\right\}.
\end{equation}
Residuals are weighted by their variances; $\log V_i$ is retained because variance depends on the parameters. We minimize this objective with bounds and multiple starts, differentiating both mean and variance. Diagnostics retain boundary solutions, solver failures, and start disagreements (Appendix~\ref{app:joint_implementation}).

Initial DI data $\mathbf Y_{\DI}$ and subsequent mode data $\mathbf Y_{\mathrm{SPADE}}$ have conditional likelihood
\begin{equation}\label{eq:adaptive_likelihood}
\begin{aligned}
&p(\mathbf Y_{\DI},\mathbf Y_{\mathrm{SPADE}}\mid\boldsymbol\theta)\\
&\quad=p(\mathbf Y_{\DI}\mid\boldsymbol\theta)\,
p(\mathbf Y_{\mathrm{SPADE}}\mid\mathbf Y_{\DI},\boldsymbol\theta).
\end{aligned}
\end{equation}
Conditioning on $\mathbf Y_{\DI}$ fixes the observed command $\mathbf u_0(\mathbf Y_{\DI})$, while true position remains unknown in the collection coefficients. Calibration supplies another factor. This represents adaptation without counting the initial DI data twice or treating their position estimate as an independent prior.

\section{Reference spectra and numerical validation}\label{app:validation}
The following checks verify optical/statistical calculations separately from the mineral receiver comparison.

\subsection{Analytical and numerical checks}
For Santamaria \emph{et al.}'s Gaussian two-source benchmark \cite{santamaria}, let $d$ be source displacement from the mode center and $\sigma$ the intensity-PSF standard deviation in the same length units. The Hermite--Gaussian mode probabilities are
\begin{equation*}
p_n=\frac{1}{n!}\exp\left(-\frac{d^2}{4\sigma^2}\right)
\left(\frac{d^2}{4\sigma^2}\right)^n.
\end{equation*}
Here $n=0,1,\ldots$ indexes mode order: a centered source occupies $n=0$, and displacement excites higher orders. Independent overlap integration agrees to approximately $1.7\times10^{-16}$ absolute error. For estimation of the target fraction in one of two spectral channels, the supplied target-to-scene photon fraction gives quadratic DI and linear resolved-mode information scaling; fitted log--log slopes are 2.0000/1.0000. This verifies theoretical probabilities and scaling, not measured counts or mineral-model exponents.

Mineral-model checks include analytical count derivatives (the Jacobian), finer source/pixel grids, independent optics, noise moments, and fixed-seed repeatability. Off-grid interpolation errors in the additional-specimen tests stay below $10^{-5}$ of target area; relative Jacobian tolerance is $10^{-4}$.

Allocation uses the simulation grid/response and per-second Poisson means with known transmission/crosstalk. QR factorization and singular value decomposition (SVD) remove column-scaled nuisance derivatives from the whitened depth derivative. Ranks and information are stable for SVD tolerances $10^{-8}$--$10^{-14}$. Centered differences agree to $2.4\times10^{-7}$ relative error; projection, scaled Schur, and uncompressed calculations agree to $3.4\times10^{-14}$ relative to DI information, and initial slopes to below $1.3\times10^{-9}$. Symmetry makes perfectly centered pure SPADE first-order position-unobservable; projection removes this redundant direction, while mixed DI supplies position information.

\subsection{Sampling plans and uncertainty intervals}
For S0--S3, target-center offsets from $(0.5L,0)$ are $(0.12L,0.08L)$, $(-0.30L,0.18L)$, $(0.24L,-0.22L)$, and $(-0.18L,-0.28L)$. S0 uses the C4 ellipse; S1 a diameter-$0.25L$ disk; S3 a $0.40L\times0.20L$ ellipse rotated $40^\circ$. S2 has radius $0.125L[1+0.2\cos(3\varphi)]$ about its center: three lobes with 20\% modulation, oriented $25^\circ$. Table~\ref{tab:pairs} gives the material pairs.

Within source square $[-2L,2L]^2$, background weights outside the target are
\begin{equation*}
\begin{aligned}
\text{S0: }&1+0.15u_x,\\
\text{S1: }&1,\\
\text{S2: }&1+0.15u_x+0.10u_y+0.12(u_x^2+u_y^2),\\
\text{S3: }&1-0.10u_x+0.15u_y+0.08(u_x^2+u_y^2).
\end{aligned}
\end{equation*}
The background is removed from the target support in every geometry.

All 24 conditions use the ten specimens withheld from design, correct references/geometry, and 500 trials per receiver: 24\,000 final estimates.

Table~\ref{tab:material_absolute} reports bias, RMSE, and sample SD (denominator $n-1$, $n=500$). Final depth/position estimates are interior; initial diagnostics remain separate. Its summary gives all conditions equal geometric weight.

Bootstrap resamples saved estimates \cite{efron1979} independently within receiver/condition. Five 100\,000-draw batches yield 500\,000 repetitions, each summary taking equal-weight geometric means. Pointwise 95\% endpoints are percentiles 2.5/97.5; Bonferroni simultaneous endpoints use $0.025/24$ and $1-0.025/24$.

Every batch identifies the same 17 pointwise and 11 Bonferroni intervals below one. Maximum batch-to-pooled Bonferroni endpoint deviation is 0.00272; increasing from 400\,000 to 500\,000 draws changes endpoints by less than 0.00041 without changing these conditions. This assesses bootstrap numerical stability, not coverage. Intervals describe finite-trial uncertainty in this fixed benchmark and exclude library/instrument uncertainty.

\begin{table*}[t]\centering\footnotesize
\caption{Absolute depth errors for the 24 supplied-spectrum conditions in Fig.~\ref{fig:external}. Each receiver has 500 trials per condition.}\label{tab:material_absolute}
\setlength{\tabcolsep}{5pt}
\begin{tabular}{@{}lrrrrrrrrl@{}}\toprule
& & \multicolumn{3}{c}{DI} & \multicolumn{3}{c}{Adaptive} & \multicolumn{2}{c}{Adaptive/DI RMSE} \\
\cmidrule(lr){3-5}\cmidrule(lr){6-8}\cmidrule(l){9-10}
Condition & True depth & Bias & SD & RMSE & Bias & SD & RMSE & Ratio & 95\% interval \\\midrule
P1--S0 & 0.36145 & -0.00064 & 0.01333 & 0.01333 & -0.00087 & 0.01055 & 0.01058 & 0.7934 & $[0.7289,0.8638]$ \\
P1--S1 & 0.36145 & -0.00054 & 0.00652 & 0.00654 & 0.00016 & 0.00542 & 0.00542 & 0.8282 & $[0.7603,0.9025]$ \\
P1--S2 & 0.36145 & -0.00078 & 0.00683 & 0.00686 & -0.00020 & 0.00555 & 0.00555 & 0.8082 & $[0.7424,0.8791]$ \\
P1--S3 & 0.36145 & -0.00010 & 0.00522 & 0.00522 & 0.00008 & 0.00404 & 0.00403 & 0.7733 & $[0.7087,0.8447]$ \\
\addlinespace[2pt]
P2--S0 & 0.44713 & -0.00060 & 0.01568 & 0.01568 & -0.00113 & 0.01348 & 0.01352 & 0.8622 & $[0.7903,0.9411]$ \\
P2--S1 & 0.44713 & 0.00038 & 0.00737 & 0.00737 & 0.00010 & 0.00640 & 0.00640 & 0.8672 & $[0.7992,0.9407]$ \\
P2--S2 & 0.44713 & 0.00030 & 0.00772 & 0.00772 & -0.00008 & 0.00738 & 0.00737 & 0.9549 & $[0.8738,1.0427]$ \\
P2--S3 & 0.44713 & -0.00027 & 0.00568 & 0.00568 & 0.00037 & 0.00509 & 0.00510 & 0.8979 & $[0.8241,0.9792]$ \\
\addlinespace[2pt]
P3--S0 & 0.44173 & 0.00095 & 0.04392 & 0.04389 & 0.00397 & 0.04543 & 0.04556 & 1.0381 & $[0.9536,1.1312]$ \\
P3--S1 & 0.44173 & 0.00025 & 0.02288 & 0.02286 & 0.00096 & 0.02240 & 0.02240 & 0.9799 & $[0.8980,1.0704]$ \\
P3--S2 & 0.44173 & -0.00197 & 0.02426 & 0.02432 & 0.00033 & 0.02451 & 0.02449 & 1.0073 & $[0.9239,1.0977]$ \\
P3--S3 & 0.44173 & -0.00090 & 0.01710 & 0.01710 & 0.00052 & 0.01662 & 0.01662 & 0.9715 & $[0.8865,1.0659]$ \\
\addlinespace[2pt]
P4--S0 & 0.37886 & 0.00062 & 0.02549 & 0.02547 & -0.00187 & 0.02423 & 0.02427 & 0.9530 & $[0.8693,1.0442]$ \\
P4--S1 & 0.37886 & -0.00042 & 0.01319 & 0.01318 & -0.00032 & 0.01126 & 0.01126 & 0.8538 & $[0.7837,0.9302]$ \\
P4--S2 & 0.37886 & 0.00018 & 0.01350 & 0.01349 & 0.00022 & 0.01315 & 0.01314 & 0.9742 & $[0.8922,1.0629]$ \\
P4--S3 & 0.37886 & -0.00039 & 0.01046 & 0.01046 & -0.00024 & 0.00933 & 0.00932 & 0.8910 & $[0.8148,0.9741]$ \\
\addlinespace[2pt]
P5--S0 & 0.32518 & 0.00006 & 0.02414 & 0.02411 & -0.00216 & 0.02077 & 0.02086 & 0.8650 & $[0.7935,0.9428]$ \\
P5--S1 & 0.32518 & 0.00025 & 0.01107 & 0.01106 & -0.00134 & 0.00976 & 0.00984 & 0.8899 & $[0.8113,0.9762]$ \\
P5--S2 & 0.32518 & 0.00028 & 0.01205 & 0.01204 & -0.00097 & 0.01073 & 0.01076 & 0.8935 & $[0.8191,0.9756]$ \\
P5--S3 & 0.32518 & -0.00027 & 0.00912 & 0.00911 & 0.00009 & 0.00796 & 0.00796 & 0.8730 & $[0.7982,0.9554]$ \\
\addlinespace[2pt]
P6--S0 & 0.22945 & -0.00354 & 0.04480 & 0.04489 & 0.00345 & 0.03498 & 0.03512 & 0.7823 & $[0.7149,0.8560]$ \\
P6--S1 & 0.22945 & -0.00012 & 0.02126 & 0.02124 & -0.00037 & 0.01701 & 0.01699 & 0.7999 & $[0.7315,0.8744]$ \\
P6--S2 & 0.22945 & -0.00072 & 0.02234 & 0.02233 & -0.00054 & 0.01961 & 0.01960 & 0.8779 & $[0.8029,0.9598]$ \\
P6--S3 & 0.22945 & -0.00089 & 0.01609 & 0.01610 & 0.00023 & 0.01353 & 0.01352 & 0.8398 & $[0.7724,0.9131]$ \\
\bottomrule\end{tabular}
\par\smallskip\raggedright
Bias, sample standard deviation (SD), and RMSE are in dimensionless band-depth units. All final depth and position boundary rates are zero for both receivers in every row; initial DI fits are reported separately in the code package. A boundary is counted within $10^{-6}$ of the fitted depth limits $[0,0.99]$ or either position limit $\pm0.6L$. The displayed intervals are pointwise percentile bootstrap intervals. The accompanying CSV also reports the Bonferroni-adjusted intervals and the boundary fractions for each condition. All estimates are retained.
\end{table*}

\section{Poisson information, pixel weighting, and time allocation}\label{app:allocation}
\subsection{Poisson information}
Let $\mathbf N$ collect the independent photon counts, with means $\mu_i(\alpha)$. The likelihood and its depth derivative are
\begin{equation}\label{eq:poisson_score}
\begin{aligned}
p(\mathbf N\mid\alpha)&=\prod_i\frac{e^{-\mu_i}\mu_i^{N_i}}{N_i!},\\
\partial_\alpha\log p&=\sum_i\frac{N_i-\mu_i}{\mu_i}\,\partial_\alpha\mu_i.
\end{aligned}
\end{equation}
Independence, zero-mean residuals, and $\operatorname{Var}(N_i)=\mu_i$ give $F_{\alpha\alpha}=\mathbb E[(\partial_\alpha\log p)^2]$ in (\ref{eq:scalar_fisher_background}); expected cross-products for two parameter derivatives give (\ref{eq:poisson_matrix}).
\subsection{Optimal local pixel weighting}
With nuisance parameters known, a weighted count has variance $\sum_iw_i^2\mu_i$ and depth derivative $\sum_iw_i\partial_\alpha\mu_i$. Cauchy--Schwarz gives
\begin{equation*}
\begin{aligned}
\left(\sum_i w_i\partial_\alpha\mu_i\right)^2
&\leq\left(\sum_i w_i^2\mu_i\right)\\
&\quad\times\left(\sum_i\frac{(\partial_\alpha\mu_i)^2}{\mu_i}\right).
\end{aligned}
\end{equation*}
For positive means and nonzero variance, divide by the first factor on the right. Equality at $w_i\propto(\partial_\alpha\mu_i)/\mu_i$ proves (\ref{eq:weights}), a local post-detection sensitivity bound.

\subsection{Exposure scaling and the allocation criterion}
Fixed receiver rates $r_i(\boldsymbol\theta)$ give $\mu_i=\tau r_i$ and
\begin{equation*}
F_{ab}(\tau)=\tau\sum_i
\frac{(\partial_{\theta_a}r_i)(\partial_{\theta_b}r_i)}{r_i}.
\end{equation*}
A known parameter-independent transmission scales the rates and information by the same factor. Independent exposures add information, giving (\ref{eq:allocation}).
The following specializes classical local optimal-design arguments \cite{pukelsheim} to the DI--SPADE matrices.
For (\ref{eq:efficient}) with positive-definite $F_{\nu\nu}$ and perturbation $v=(1,v_\nu\tr)\tr$, completing the square gives
\begin{equation}\label{eq:complete_square}
\begin{aligned}
v\tr Fv
&=F_{\alpha\alpha}+2F_{\alpha\nu}v_\nu
  +v_\nu\tr F_{\nu\nu}v_\nu\\
&=I_\eff
  +(v_\nu+F_{\nu\nu}^{-1}F_{\nu\alpha})\tr\\
&\qquad\times F_{\nu\nu}
  (v_\nu+F_{\nu\nu}^{-1}F_{\nu\alpha}).
\end{aligned}
\end{equation}
The last term is nonnegative and vanishes at $v_\nu=-F_{\nu\nu}^{-1}F_{\nu\alpha}$; hence
\begin{equation}\label{eq:variational}
I_\eff(F)=\min_{v_\alpha=1}v\tr Fv.
\end{equation}
Fixing $v_\alpha=1$ makes $v_\nu$ a nuisance change per unit depth change; the minimizer gives (\ref{eq:di_direction}).

Each $v\tr F(t)v$ is affine in $t$, so its minimum is concave. Let $v(t)$ minimize it and write $I_\eff(t)=I_\eff[F(t)]$. Differentiating,
\begin{equation*}
\begin{aligned}
I_\eff'(t)={}&v(t)\tr F'(t)v(t)\\
&+2v'(t)\tr F(t)v(t).
\end{aligned}
\end{equation*}
Stationarity makes the nuisance components of $F(t)v(t)$ zero; $v_\alpha(t)=1$ gives $v'_\alpha(t)=0$. Thus the second term vanishes. With $F'(t)=\eta F^{(\mathrm{SPADE})}-F^{(\DI)}$ and $v(0)=v_{\DI}$,
\begin{equation}\label{eq:allocation_derivative}
I_\eff'(0)=\eta v_{\DI}\tr F^{(\mathrm{SPADE})}v_{\DI}-I_{\DI}.
\end{equation}
A positive right derivative improves on DI for sufficiently small $t>0$. Otherwise concavity gives
\begin{equation*}
I_\eff(t)\leq I_{\DI}+tI_\eff'(0)\leq I_{\DI}.
\end{equation*}
This proves (\ref{eq:derivative}) under fixed-response Poisson measurements. Positive-definite $F_{\nu\nu}^{(\DI)}$ keeps the mixed nuisance block invertible for $0\leq t<1$. At a singular pure-SPADE endpoint, the variational comparison still exists, but an inverse-matrix variance bound requires local parameter identifiability. Parameter-boundary effects and nonlinear acquisition costs are outside this criterion.

\begin{table*}[!t]\centering\footnotesize
\caption{Independent confirmation: mean depth error, sample standard deviation (SD), and RMSE in absorption-depth units. Each row has 400 trials per receiver. The last two rows use the original depth and the reference errors described in Fig.~\ref{fig:depth_confirmation}.}\label{tab:independent_depth}
\setlength{\tabcolsep}{5pt}
\begin{tabular}{lrrrrrr}\toprule
& \multicolumn{3}{c}{DI} & \multicolumn{3}{c}{Adaptive} \\
Condition & Mean error & SD & RMSE & Mean error & SD & RMSE \\\midrule
$\alpha=0.10$ & -0.00029 & 0.02500 & 0.02497 & 0.00080 & 0.02095 & 0.02094 \\
$\alpha=0.20$ & -0.00090 & 0.02265 & 0.02264 & -0.00331 & 0.01915 & 0.01941 \\
$\alpha=0.25053$ & -0.00109 & 0.02388 & 0.02387 & -0.00248 & 0.01761 & 0.01776 \\
$\alpha=0.30$ & -0.00092 & 0.02231 & 0.02230 & -0.00052 & 0.01847 & 0.01846 \\
$\alpha=0.40$ & 0.00031 & 0.02155 & 0.02152 & -0.00124 & 0.01703 & 0.01706 \\
Background-only error & 0.00252 & 0.02271 & 0.02282 & -0.00124 & 0.01850 & 0.01851 \\
Target and background error & 0.00028 & 0.02353 & 0.02350 & -0.00112 & 0.01917 & 0.01918 \\
\bottomrule\end{tabular}\end{table*}

\begin{table*}[t]\centering\footnotesize
\caption{Spatial-background stress test of the frozen inference pipeline. All nine predeclared conditions are retained, with 400 trials per receiver. The uniform control uses the same mean quartz fraction as its gradient counterpart. Intervals are pointwise 95\% independent-arm bootstrap intervals. Strong-gradient rows include local-optimization ambiguity and parameter-boundary solutions; their ratios are pipeline diagnostics, not established physical precision boundaries.}\label{tab:spatial_background}
\setlength{\tabcolsep}{3.5pt}
\begin{tabular}{lrrrrrrl}\toprule
& \multicolumn{3}{c}{DI} & \multicolumn{3}{c}{Adaptive} & Adaptive/DI \\
Condition & Mean error & SD & RMSE & Mean error & SD & RMSE & Ratio [95\% interval] \\\midrule
Original C4 & -0.00044 & 0.02196 & 0.02194 & -0.00273 & 0.01863 & 0.01880 & 0.857 [0.773, 0.950] \\
Uniform, $a=0.02$ & 0.00079 & 0.02243 & 0.02241 & -0.00276 & 0.01847 & 0.01866 & 0.832 [0.752, 0.921] \\
Gradient, $a=0.02$ & -0.03035 & 0.02218 & 0.03758 & -0.02291 & 0.01850 & 0.02943 & 0.783 [0.724, 0.847] \\
Uniform, $a=0.05$ & 0.00051 & 0.02332 & 0.02329 & -0.00354 & 0.01902 & 0.01932 & 0.829 [0.751, 0.914] \\
Gradient, $a=0.05$ & -0.04330 & 0.02886 & 0.05202 & -0.06189 & 0.07582 & 0.09780 & 1.880 [1.695, 2.070] \\
Uniform, $a=0.1$ & 0.00058 & 0.02317 & 0.02315 & -0.00338 & 0.01853 & 0.01882 & 0.813 [0.738, 0.897] \\
Gradient, $a=0.1$ & 0.06941 & 0.04529 & 0.08285 & 0.10140 & 0.09248 & 0.13716 & 1.656 [1.545, 1.770] \\
Uniform, $a=0.2$ & 0.00147 & 0.02159 & 0.02161 & -0.00270 & 0.01877 & 0.01894 & 0.876 [0.792, 0.969] \\
Gradient, $a=0.2$ & 0.22577 & 0.02194 & 0.22683 & 0.11076 & 0.11049 & 0.15635 & 0.689 [0.619, 0.756] \\
\bottomrule\end{tabular}\end{table*}

\section{Spectral estimation and confirmation details}\label{app:references}\label{sec:uncertainty}
\subsection{Target continuum and background parameterization}\label{app:joint_implementation}
Target continuum slope $s_T$ and curvature $q_T$ are dimensionless tilt/bending coefficients, not wavelength derivatives; $\beta_T$ scales brightness. To compare wavelength windows, define
\begin{equation*}
\xi_\ell=\frac{2(\lambda_\ell-\bar\lambda)}{\lambda_{\max}-\lambda_{\min}}.
\end{equation*}
Here $\bar\lambda$ is the mean sampled wavelength and $\lambda_{\min},\lambda_{\max}$ the endpoints, so spectral coordinate $\xi_\ell$ spans approximately $[-1,1]$. The extended model is
\begin{equation}\label{eq:continuum_extension}
\begin{aligned}
R_{T\ell}={}&\beta_T C_\ell(1-\alpha h_\ell)\\
&\times(1+s_T\xi_\ell)
\left[1+q_T(\xi_\ell^2-\tfrac13)\right].
\end{aligned}
\end{equation}
Positive $s_T$ raises the long-wavelength side; positive $q_T$ raises both ends relative to the middle. Subtracting $1/3$, the continuous uniform mean of $\xi^2$ on $[-1,1]$, removes the mean quadratic perturbation, without implying coefficient independence after sampling/filtering.

The main acquisition fits $s_T\in[-0.05,0.05]$ with $q_T=0$; additional-material tests fix both and supply the background shape. Depth always refers to the modeled continuum and increasing depth lowers flux.

For optimization, the 25 logarithmic background corrections are parameterized as
\begin{equation}\label{eq:background_basis}
\boldsymbol\gamma=U\mathbf c,
\end{equation}
where $U$ is the full $25\times25$ orthonormal discrete cosine transform (DCT) matrix and each entry of $\mathbf c$ is bounded by $\pm3$. All columns are retained: this is a coordinate change without smoothing, already encompassing background brightness/slope.

Initial and full DI fits share 30 parameters: depth, target brightness/slope, position, and 25 background values. The adaptive final fit adds $\eta,\epsilon$ and two calibration-source intensities. Bounded damped Fisher scoring uses multiple position starts and L-BFGS-B fallback, retaining derivatives of mean and variance. Confirmation records convergence and boundary diagnostics.

\subsection{Background reference and target-continuum information}\label{app:reference_invariance}
For positive background references $a,b$ on the same grid, hold the target template, optics, calibration, and recorded mode center fixed and set
\begin{equation}\label{eq:reference_shift}
\gamma_\ell^{(b)}
=\gamma_\ell^{(a)}
+\log\frac{R_{B\ell}^{\mathrm{ref},(a)}}{R_{B\ell}^{\mathrm{ref},(b)}}.
\end{equation}
Substitution in (\ref{eq:free_background}) gives
\begin{equation}
R_{B\ell}^{\mathrm{ref},(b)}e^{\gamma_\ell^{(b)}}
=R_{B\ell}^{\mathrm{ref},(a)}e^{\gamma_\ell^{(a)}}.
\end{equation}
Thus background spectra, means, variances, and objective (\ref{eq:fit}) are unchanged. In the full DCT basis this is a translation by $U^\mathsf{T}$ times the reference log-ratio vector. Equivalence holds in the unrestricted positive model or when both coefficient boxes contain the solution; arbitrary translations need not preserve finite bounds.

Replacing only the background reference tests coordinate and optimization stability. Changes in target template or spatial composition alter the physical model. Section~\ref{sec:scope} reports target-curvature information.

Curvature diagnostics use seed 202610220734 and Table~\ref{tab:sensitivity}'s C4 settings, generating continua, 400 trials per receiver, and 10 s accounting. Both fit 25 background values, with $q_T=0$ or $q_T\in[-0.03,0.03]$. The bounds are illustrative, not independently measured priors. No final depth boundary or solver failure occurs. Independent-arm bootstrap (20\,000 draws) gives $[0.7864,0.9567]$/$[0.8329,0.9985]$ for omitted/fitted curvature. These intervals condition on the fixed cases and bounds; retained curvature-bound activity precludes general continuum robustness. The runs precede the frozen confirmation; S2 links their complete records.

\subsection{Reference perturbations in the confirmation}
Reference-error tests keep true C4 spectra fixed and perturb estimator inputs at native USGS wavelengths. Zero-mean Gaussian errors have recorded standard errors of the mean (SEM) as marginal deviations and correlation
\begin{equation}
\operatorname{Corr}(\delta R_\ell,\delta R_m)
=\exp\!\left[-\frac{(\lambda_\ell-\lambda_m)^2}{2(12\ {\rm nm})^2}\right],
\end{equation}
Here $\delta R_\ell$ denotes reference error. Perturbed spectra are interpolated to the model grid, with positive reflectances. One test perturbs background only; the other both references plus a common uniform offset within $\pm0.5$ nm. Target continuum/profile are reconstructed from the perturbed reference without supplying realized errors. These are recorded-uncertainty-scaled perturbations, not a field-accuracy specification.

\subsection{Independent confirmation of the frozen estimator}\label{app:independent_depth}
Confirmation uses science/reference seeds 202610221145/202610221146 and hash-verified frozen code. Only depth varies under Sec.~\ref{sec:depth_confirmation}'s settings. Exact-reference arms resample independent trials; perturbed-reference arms resample paired groups of ten observations.

Five 100\,000-draw batches use Bonferroni tails $0.025/7$ and $1-0.025/7$; all identify seven adjusted intervals below one. All 5600 final and 2800 initial estimates are retained, with no reported solver failures. Forty-two predefined extra-start checks change depth by at most $4.9\times10^{-7}$ and improve the objective by less than $1.9\times10^{-9}$.

Table~\ref{tab:independent_depth} reports bias, SD, and RMSE. At depth 0.10, 22 initial depth and four initial position estimates reach bounds; one final adaptive position estimate does too. No final depth/background coefficient reaches a bound. These trials remain included despite poorer initial localization.

\section{Spatial-background model diagnostic}\label{app:spatial_background}
To test spatially shared background shape, we mix dolomite with valid USGS quartz HS32 reflectance covering the C4 window \cite{usgs}. Background remains excluded from the target. The quartz fraction is

\begin{equation}
\zeta(\mathbf u)=a\,\frac{u_x+2}{4},
\qquad -2\leq u_x\leq2,
\end{equation}
Here $a$ is maximum quartz fraction, with dolomite fraction $1-\zeta$. Their linear spectral mixture is multiplied by the original background brightness weight. Predeclared $a=0,0.02,0.05,0.10,0.20$ each have a uniform control at the target-excluded brightness-weighted mean, approximately $0.549823a$. Paired controls/gradients therefore share total flux at every wavelength before spatial selection.

The nine conditions retain the frozen shared-spectrum fitter and 10 s budget, with no supplied quartz template, mixture fraction, or gradient. Each uses 400 acquisitions per receiver, independent science seed 202610220921, separate receiver/stage streams, and 100\,000-draw independent-arm bootstrap intervals.

The generator recovers C4 at $a=0$ and represents uniform mixtures to relative mean error below $8\times10^{-16}$. Independent surface/pixel quadrature verifies both responses. Table~\ref{tab:spatial_background} retains all conditions: uniform controls have small errors, whereas spatial gradients create bias despite freely fitted spectral values.

All frozen fits terminate numerically, but extra starts find lower objectives in some $a\geq0.05$ cases. At $a=0.10,0.20$, every final position reaches a boundary in both receivers. Convergence alone therefore does not establish the best likelihood solution under this spatial mismatch. All boundary and extra-start diagnostics are archived.

\phantomsection\label{sec:references}
\bibliographystyle{IEEEtran}
\bibliography{refs}

@article{kruse,
 author={Kruse, Fred A. and Bedell, Richard L. and Taranik, James V. and Peppin, William A. and Weatherbee, Oliver and Calvin, Wendy M.},
 title={Mapping alteration minerals at prospect, outcrop and drill core scales using imaging spectrometry},
 journal={Int. J. Remote Sens.}, year={2012}, volume={33}, number={6}, pages={1780--1798}, doi={10.1080/01431161.2011.600350}}

@article{swayze,
 author={Swayze, Gregg A. and Clark, Roger N. and Goetz, Alexander F. H. and Chrien, Thomas G. and Gorelick, Noel S.},
 title={Effects of spectrometer band pass, sampling, and signal-to-noise ratio on spectral identification using the {Tetracorder} algorithm},
 journal={J. Geophys. Res., Planets}, year={2003}, volume={108}, number={E9}, pages={5105}, doi={10.1029/2002JE001975}}

@article{tsang,
 author={Tsang, Mankei and Nair, Ranjith and Lu, Xiao-Ming},
 title={Quantum Theory of Superresolution for Two Incoherent Optical Point Sources},
 journal={Phys. Rev. X}, year={2016}, volume={6}, pages={031033}, doi={10.1103/PhysRevX.6.031033}}

@article{huang,
 author={Huang, Zixin and Schwab, Christian and Lupo, Cosmo},
 title={Ultimate limits of exoplanet spectroscopy: A quantum approach},
 journal={Phys. Rev. A}, year={2023}, volume={107}, pages={022409}, doi={10.1103/PhysRevA.107.022409}}

@article{santamaria,
 author={Santamaria, Luigi and Sgobba, Fabrizio and Pallotti, Deborah and Lupo, Cosmo},
 title={Single-photon super-resolved spectroscopy from spatial-mode demultiplexing},
 journal={Photonics Res.}, year={2025}, volume={13}, number={4}, pages={865--874}, doi={10.1364/PRJ.544197}}

@article{grace,
 author={Grace, Michael R. and Dutton, Zachary and Ashok, Amit and Guha, Saikat},
 title={Approaching quantum-limited imaging resolution without prior knowledge of the object location},
 journal={J. Opt. Soc. Am. A}, year={2020}, volume={37}, number={8}, pages={1288--1299}, doi={10.1364/JOSAA.392116}}

@article{prasad,
 author={Prasad, Sudhakar},
 title={Quantum Limits on Localizing Point Objects against a Uniformly Bright Disk},
 journal={Phys. Rev. A}, year={2023}, volume={107}, pages={032427}, doi={10.1103/PhysRevA.107.032427}}

@misc{h2rg,
 author={{Teledyne Imaging Sensors}},
 title={{H2RG} Visible and Infrared Focal Plane Array}, year={2022}, month=feb,
 note={Specification sheet TSI-0855; 2.5-micrometer array, CDS noise at 100 kHz},
 url={https://www.teledyne-si.com/en-us/Products-and-Services_/Documents/Infrared%20and%20Visible%20FPAs/TSI-0855%20H2RG%20Brochure-25Feb2022.pdf}}

@misc{specim,
 author={{Specim}}, title={Spectral Camera {SWIR}}, note={Specification sheet CSWI-D-3-15},
 url={https://qd-europe.com/fileadmin/Mediapool/products/Specim/_pdf/Spectral_cameras_SWIR.pdf}}

@techreport{usgs,
 author={Clark, Roger N. and others}, title={{USGS Digital Spectral Library splib05a}},
 institution={U.S. Geological Survey}, type={Open-File Report}, number={2003-395}, year={2003}, doi={10.3133/ofr03395}}

@article{ansari,
 author={Ansari, Vahid and Brecht, Benjamin and Gil-Lopez, Jano and Donohue, John M. and {\v{R}eh\'{a}\v{c}ek}, Jaroslav and Hradil, Zden{\v{e}}k and S{\'{a}}nchez-Soto, Luis L. and Silberhorn, Christine},
 title={Achieving the Ultimate Quantum Timing Resolution},
 journal={PRX Quantum}, year={2021}, volume={2}, pages={010301}, doi={10.1103/PRXQuantum.2.010301}}

@article{clarkroush,
 author={Clark, Roger N. and Roush, Ted L.},
 title={Reflectance spectroscopy: Quantitative analysis techniques for remote sensing applications},
 journal={J. Geophys. Res., Solid Earth}, year={1984}, volume={89}, number={B7}, pages={6329--6340}, doi={10.1029/JB089iB07p06329}}

@article{manolakis,
 author={Manolakis, Dimitris and Siracusa, Christina and Shaw, Gary},
 title={Hyperspectral subpixel target detection using the linear mixing model},
 journal={IEEE Trans. Geosci. Remote Sens.}, year={2001}, volume={39}, number={7}, pages={1392--1409}, doi={10.1109/36.934072}}

@article{kerekes,
 author={Kerekes, John P. and Baum, Jerrold E.},
 title={Spectral imaging system analytical model for subpixel object detection},
 journal={IEEE Trans. Geosci. Remote Sens.}, year={2002}, volume={40}, number={5}, pages={1088--1101}, doi={10.1109/TGRS.2002.1010896}}

@article{greenberger,
 author={Greenberger, Rebecca N. and Mustard, John F. and Ehlmann, Bethany L. and Blaney, Diana L. and Cloutis, Edward A. and Wilson, Janette H. and Green, Robert O. and Fraeman, Abigail A.},
 title={Imaging spectroscopy of geological samples and outcrops: Novel insights from microns to meters},
 journal={GSA Today}, year={2015}, volume={25}, number={12}, pages={4--10}, doi={10.1130/GSATG252A.1}}

@article{koerting,
 author={Koerting, Friederike and Koellner, Nicole and Kuras, Agnieszka and Boesche, Nina Kristin and Rogass, Christian and Mielke, Christian and Elger, Kirsten and Altenberger, Uwe},
 title={A solar optical hyperspectral library of rare-earth-bearing minerals, rare-earth oxide powders, copper-bearing minerals and Apliki mine surface samples},
 journal={Earth Syst. Sci. Data}, year={2021}, volume={13}, pages={923--942}, doi={10.5194/essd-13-923-2021}}

@misc{emitdepth,
 author={Green, Robert},
 title={{EMIT L2B} Estimated Mineral Identification and Band Depth and Uncertainty 60 m {V001}},
 year={2023},
 howpublished={NASA Land Processes Distributed Active Archive Center},
 note={Data set},
 doi={10.5067/EMIT/EMITL2BMIN.001}}

@book{pukelsheim,
 author={Pukelsheim, Friedrich},
 title={Optimal Design of Experiments},
 series={Classics in Applied Mathematics}, volume={50},
 publisher={Society for Industrial and Applied Mathematics},
 address={Philadelphia, PA, USA}, year={2006},
 doi={10.1137/1.9780898719109}}

@article{asadzadeh2016,
 author={Asadzadeh, Saeid and de Souza Filho, Carlos Roberto},
 title={Iterative Curve Fitting: A Robust Technique to Estimate the Wavelength Position and Depth of Absorption Features From Spectral Data},
 journal={IEEE Trans. Geosci. Remote Sens.},
 year={2016}, volume={54}, number={10}, pages={5964--5974},
 doi={10.1109/TGRS.2016.2577621}}

@article{liu2023,
 author={Liu, Wanyue and Hecker, Christoph and van Ruitenbeek, Frank J. A. and van Eijndthoven, Wijnand and Cheng, Qiuming},
 title={Identifying and Quantifying Carbonate Minerals in Quartz--Illite--Muscovite-Dominated Reservoir Rocks With {SWIR} and {LWIR} Spectroscopies},
 journal={IEEE Trans. Geosci. Remote Sens.},
 year={2023}, volume={61}, pages={5003813},
 doi={10.1109/TGRS.2023.3298902}}

@article{goetz1985,
 author={Goetz, Alexander F. H. and Vane, Gregg and Solomon, Jerry E. and Rock, Barrett N.},
 title={Imaging Spectrometry for {Earth} Remote Sensing},
 journal={Science},
 year={1985},
 volume={228},
 number={4704},
 pages={1147--1153},
 doi={10.1126/science.228.4704.1147}}

@article{hunt1977,
 author={Hunt, G. R.},
 title={Spectral signatures of particulate minerals in the visible and near infrared},
 journal={Geophysics},
 year={1977},
 volume={42},
 number={3},
 pages={501--513},
 doi={10.1190/1.1440721}}

@article{clark1990,
 author={Clark, Roger N. and King, Trude V. V. and Klejwa, Matthew and Swayze, Gregg A. and Vergo, Norma},
 title={High spectral resolution reflectance spectroscopy of minerals},
 journal={J. Geophys. Res., Solid Earth},
 year={1990},
 volume={95},
 number={B8},
 pages={12653--12680},
 doi={10.1029/jb095ib08p12653}}

@article{sunshine1990,
 author={Sunshine, Jessica M. and Pieters, Carle M. and Pratt, Stephen F.},
 title={Deconvolution of mineral absorption bands: An improved approach},
 journal={J. Geophys. Res., Solid Earth},
 year={1990},
 volume={95},
 number={B5},
 pages={6955--6966},
 doi={10.1029/jb095ib05p06955}}

@article{hapke1981,
 author={Hapke, Bruce},
 title={Bidirectional reflectance spectroscopy: 1. Theory},
 journal={J. Geophys. Res., Solid Earth},
 year={1981},
 volume={86},
 number={B4},
 pages={3039--3054},
 doi={10.1029/jb086ib04p03039}}

@article{vandermeer2012,
 author={van der Meer, Freek D. and van der Werff, Harald M.A. and van Ruitenbeek, Frank J.A. and Hecker, Chris A. and Bakker, Wim H. and Noomen, Marleen F. and van der Meijde, Mark and Carranza, E. John M. and de Smeth, J. Boudewijn and Woldai, Tsehaie},
 title={Multi- and hyperspectral geologic remote sensing: A review},
 journal={Int. J. Appl. Earth Obs. Geoinf.},
 year={2012},
 volume={14},
 number={1},
 pages={112--128},
 doi={10.1016/j.jag.2011.08.002}}

@article{bioucas2012,
 author={Bioucas-Dias, José M. and Plaza, Antonio and Dobigeon, Nicolas and Parente, Mario and Du, Qian and Gader, Paul and Chanussot, Jocelyn},
 title={Hyperspectral Unmixing Overview: Geometrical, Statistical, and Sparse Regression-Based Approaches},
 journal={IEEE J. Sel. Topics Appl. Earth Observ. Remote Sens.},
 year={2012},
 volume={5},
 number={2},
 pages={354--379},
 doi={10.1109/jstars.2012.2194696}}

@article{nair2016,
 author={Nair, Ranjith and Tsang, Mankei},
 title={Far-Field Superresolution of Thermal Electromagnetic Sources at the Quantum Limit},
 journal={Phys. Rev. Lett.},
 year={2016},
 volume={117},
 number={19},
 pages={190801},
 doi={10.1103/physrevlett.117.190801}}

@article{paur2016,
 author={Paúr, Martin and Stoklasa, Bohumil and Hradil, Zdenek and Sánchez-Soto, Luis L. and Rehacek, Jaroslav},
 title={Achieving the ultimate optical resolution},
 journal={Optica},
 year={2016},
 volume={3},
 number={10},
 pages={1144--1147},
 doi={10.1364/optica.3.001144}}

@article{tham2017,
 author={Tham, Weng-Kian and Ferretti, Hugo and Steinberg, Aephraim M.},
 title={Beating {Rayleigh}'s Curse by Imaging Using Phase Information},
 journal={Phys. Rev. Lett.},
 year={2017},
 volume={118},
 number={7},
 pages={070801},
 doi={10.1103/physrevlett.118.070801}}

@article{tang2016,
 author={Tang, Zong Sheng and Durak, Kadir and Ling, Alexander},
 title={Fault-tolerant and finite-error localization for point emitters within the diffraction limit},
 journal={Opt. Express},
 year={2016},
 volume={24},
 number={19},
 pages={22004--22012},
 doi={10.1364/oe.24.022004}}

@article{tan2023,
 author={Tan, Xiao-Jie and Qi, Luo and Chen, Lianwei and Danner, Aaron J. and Kanchanawong, Pakorn and Tsang, Mankei},
 title={Quantum-inspired superresolution for incoherent imaging},
 journal={Optica},
 year={2023},
 volume={10},
 number={9},
 pages={1189--1194},
 doi={10.1364/optica.493227}}

@article{boucher2020,
 author={Boucher, Pauline and Fabre, Claude and Labroille, Guillaume and Treps, Nicolas},
 title={Spatial optical mode demultiplexing as a practical tool for optimal transverse distance estimation},
 journal={Optica},
 year={2020},
 volume={7},
 number={11},
 pages={1621--1626},
 doi={10.1364/optica.404746}}

@article{morizur2010,
 author={Morizur, Jean-François and Nicholls, Lachlan and Jian, Pu and Armstrong, Seiji and Treps, Nicolas and Hage, Boris and Hsu, Magnus and Bowen, Warwick and Janousek, Jiri and Bachor, Hans-A.},
 title={Programmable unitary spatial mode manipulation},
 journal={J. Opt. Soc. Am. A},
 year={2010},
 volume={27},
 number={11},
 pages={2524--2531},
 doi={10.1364/josaa.27.002524}}

@article{fontaine2019,
 author={Fontaine, Nicolas K. and Ryf, Roland and Chen, Haoshuo and Neilson, David T. and Kim, Kwangwoong and Carpenter, Joel},
 title={{Laguerre--Gaussian} mode sorter},
 journal={Nat. Commun.},
 year={2019},
 volume={10},
 pages={1865},
 doi={10.1038/s41467-019-09840-4}}

@article{gessner2020,
 author={Gessner, Manuel and Fabre, Claude and Treps, Nicolas},
 title={Superresolution Limits from Measurement Crosstalk},
 journal={Phys. Rev. Lett.},
 year={2020},
 volume={125},
 number={10},
 pages={100501},
 doi={10.1103/physrevlett.125.100501}}

@article{bonsma2019,
 author={Bonsma-Fisher, Kent A G and Tham, Weng-Kian and Ferretti, Hugo and Steinberg, Aephraim M},
 title={Realistic sub-{Rayleigh} imaging with phase-sensitive measurements},
 journal={New J. Phys.},
 year={2019},
 volume={21},
 number={9},
 pages={093010},
 doi={10.1088/1367-2630/ab3d97}}

@article{wallis2025,
 author={Wallis, John S. and Gozzard, David R. and Frost, Alex M. and Collier, Joshua J. and Maron, Nicolas and Dix-Matthews, Benjamin P.},
 title={Spatial mode demultiplexing for super-resolved source parameter estimation},
 journal={Opt. Express},
 year={2025},
 volume={33},
 number={16},
 pages={34651--34662},
 doi={10.1364/oe.563503}}

@article{elfving1952,
 author={Elfving, G.},
 title={Optimum Allocation in Linear Regression Theory},
 journal={Ann. Math. Statist.},
 year={1952},
 volume={23},
 number={2},
 pages={255--262},
 doi={10.1214/aoms/1177729442}}

@article{efron1979,
 author={Efron, B.},
 title={Bootstrap Methods: Another Look at the Jackknife},
 journal={Ann. Statist.},
 year={1979},
 volume={7},
 number={1},
 pages={1--26},
 doi={10.1214/aos/1176344552}}
\label{sec:references_end}
\end{document}